\documentclass[english,aps,prc,nofootinbib,superscriptaddress,twocolumn,letter]{revtex4}
\usepackage[latin9]{inputenc}
\usepackage{hyperref}
\hypersetup{
  colorlinks=true,        % false: boxed links; true: colored links
  linkcolor=blue,         % color of internal links
  citecolor=cyan,         % color of links to bibliography
}
\usepackage{breakurl}
\usepackage{graphicx}
\usepackage{epsf}
\usepackage{epsfig}
\usepackage{amssymb,amsmath}
\usepackage[usenames]{color}
\usepackage{amssymb}
\usepackage{times}
\usepackage{comment}

\newcommand\snn{\sqrt{s_\text{NN}}}

\allowdisplaybreaks



\begin{document}

\preprint{This line only printed with preprint option}

\title{Modified Trento initial condition and its impact on collective flows and global polarization in Cu+Au collisions}

\author{Ze-Fang Jiang}
\email{jiangzf@mails.ccnu.edu.cn}
\affiliation{Department of Physics and Electronic-Information Engineering, Hubei Engineering University, Xiaogan, Hubei, 432000, China}
\affiliation{Institute of Particle Physics and Key Laboratory of Quark and Lepton Physics (MOE), Central China Normal University, Wuhan, Hubei, 430079, China}

\author{Shanshan Cao}
\email{shanshan.cao@sdu.edu.cn}
\affiliation{Institute of Frontier and Interdisciplinary Science, Shandong University, Qingdao, Shandong, 266237, China}

\author{Ben-Wei Zhang}
\email{bwzhang@mail.ccnu.edu.cn}
\affiliation{Institute of Particle Physics and Key Laboratory of Quark and Lepton Physics (MOE), Central China Normal University, Wuhan, Hubei, 430079, China}

\begin{abstract}
Collective flow coefficients and spin polarization are valuable probes of the geometry and flow velocity field of the quark-gluon plasma (QGP) produced in relativistic heavy-ion collisions. Using a modified TRENTo initial condition coupled to a (3+1)-dimensional (D) viscous hydrodynamic model CLVisc, we study the directed flow and elliptic flow coefficients of hadrons, together with the global polarization of $\Lambda$ and $\bar{\Lambda}$ hyperons in asymmetric Cu+Au collisions. We extend the 2D TRENTo model to the 3D space, and find that the initial tilted geometry of the QGP fireball with respect to the longitudinal direction leads to a decrease of directed flow from positive to negative values with increasing pseudorapidity, an enhancement of elliptic flow at forward and backward pseudorapidities, and a non-monotonic dependence of global polarization on the transverse momentum of hyperons. The initial longitudinal flow velocity gradient further enhances the values of directed flow and global polarization. Our model calculation provides a satisfactory description of the rapidity, transverse momentum, and centrality dependences of the directed flow of charged hadrons in Cu+Au collisions for the first time, and proposes that such asymmetric heavy-ion collisions create a better environment for studying the initial tilted geometry and longitudinal flow field of the QGP than symmetric Au+Au collisions do.

\end{abstract}
%\pacs{20.24, 20.25}
\maketitle
\date{\today}

\section{Introduction}
\label{emsection1}

A hot and dense quantum chromodynamics (QCD) matter, known as quark-gluon plasma (QGP), has been extensively studied in high-energy nuclear collisions at the Relativistic Heavy-Ion Collider (RHIC) and the Large Hadron Collider (LHC)~\cite{Busza:2018rrf}. 
One of the smoking gun signals of the strongly coupled nature of the QGP is the large values of the collective flow coefficients of hadrons emitted from the QGP~\cite{STAR:2003wqp,ALICE:2010suc,ATLAS:2014txd,CMS:2012zex}, which are defined as the Fourier coefficients of the azimuthal angle distribution of the hadron momenta~\cite{Voloshin:2008dg,Bilandzic:2010jr} and have been successfully described by hydrodynamic models~\cite{Huovinen:2001cy,Kolb:2003dz,Gale:2013da,Shen:2020mgh}. Small shear viscosity, close to the minimum bound obtained with an infinitely strong coupling~\cite{Policastro:2001yc,Kovtun:2004de}, has been extracted from comparisons between hydrodynamic calculations and experimental data~\cite{Bernhard:2019bmu,JETSCAPE:2020shq}, indicating the QGP is a strongly interacting matter that is closest to a perfect fluid man has ever achieved in laboratory. Apart from transport properties of the QGP, the second and higher order collective flow coefficients have also been used to constrain the initial condition of heavy-ion collisions, such as the initial energy density distribution of the QGP and its quantum fluctuations~\cite{Schenke:2011bn,Gale:2012rq,Heinz:2013th,Chen:2024xbi}, shapes of different species of nuclei~\cite{Zhang:2021kxj,Jia:2021qyu,Bally:2022vgo}, and even the size of a nucleon bound inside a nucleus~\cite{Giacalone:2021clp} and the sub-nucleon structure~\cite{Zhu:2024tns}.

Although hydrodynamic description of the second and higher order collective flows has reached pretty high precision, study on the first order collective flow, the directed flow ($v_1$), is still at an early stage. Considerable efforts have been devoted to understanding the production mechanisms of $v_1$ and its evolution through the QGP expansion. For example, it has been proposed that a tilted geometry of the QGP fireball in the reaction plane with respect to the longitudinal direction is favored by the experimental data on the rapidity odd $v_1$ of hadrons~\cite{Bozek:2010bi,Jiang:2021ajc} and a tilted distribution of the net baryon density distribution further helps explain the splitting of $v_1$ between baryons and anti-baryons~\cite{Bozek:2022svy,Parida:2022zse,Jiang:2023fad}. The longitudinal flow velocity gradient in the initial state, which is beyond the Bjorken approximation, is also found important for describing the hadron $v_1$~\cite{Shen:2020jwv}, especially if one wants to describe the phenomenology of directed flow and global polarization of hyperons within the same hydrodynamic framework~\cite{Ryu:2021lnx,Jiang:2023vxp}. Effects of the initial state fluctuations~\cite{Csernai:2011gg,Retinskaya:2012ky} and the equation of state of the QGP~\cite{Ivanov:2016sqy,Nara:2016hbg,Du:2022yok} on $v_1$, and the dependences of $v_1$ on the electric charge and strangeness of constituent quarks of hadrons~\cite{Nayak:2023ofv} have also been explored. Nevertheless, a simultaneous description of the experimental data on the directed flow coefficients of different species of hadrons and at different rapidity regions and different beam energies of heavy-ion collisions~\cite{Selyuzhenkov:2011zj,ATLAS:2012at,STAR:2016cio,STAR:2022fnj,ALICE:2013xri,STAR:2017okv,STAR:2023jdd,STAR:2024ujm} still remains challenging. Measurements on $v_1$ in Cu+Au collisions at STAR~\cite{STAR:2017ykf} provides an additional opportunity for constraining the dependences of $v_1$ on the geometry and flow velocity profiles of the QGP, considering the initial condition of the asymmetric Cu+Au collisions is different from that of the symmetric Au+Au or Pb+Pb collisions. Predictions on $v_1$ based on hydrodynamic calculations~\cite{Bozek:2012hy,Nakamura:2022idq} and transport models~\cite{Chen:2005zy,Voronyuk:2014rna} for Cu+Au collisions exist in literature, but so far, a satisfactory description of the experimental data has not been achieved. The main purpose of the present work is to develop an initial condition model for understanding various origins of the hadron $v_1$ in asymmetric heavy-ion collisions.

\begin{figure}[tp!]
\begin{center}
\includegraphics[width=0.90\linewidth]{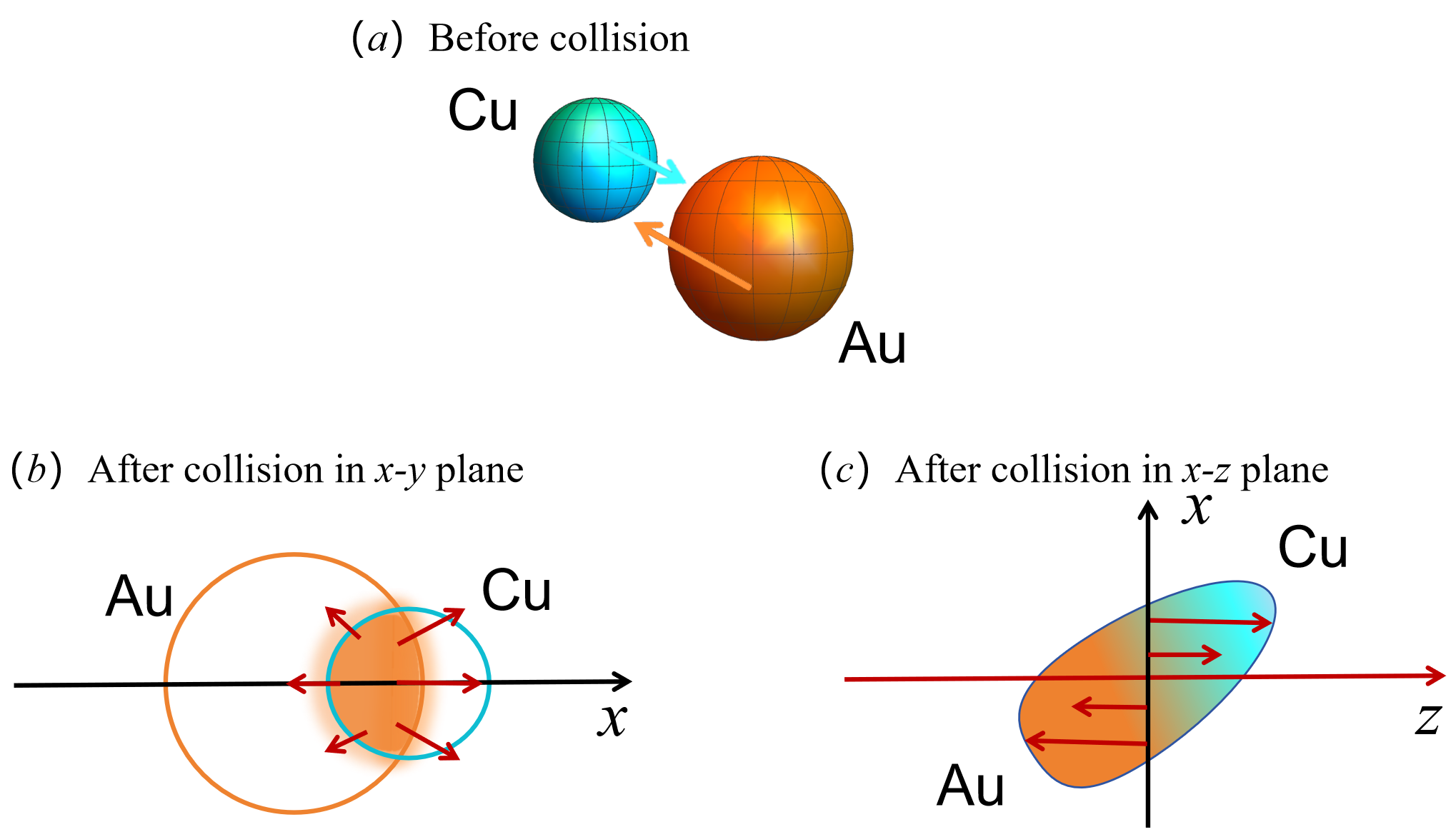} 
\end{center}
\caption{(Color online) Cartoons of non-central Cu+Au collisions, in which a Cu nucleus traveling along the $+\hat{z}$ direction on the $+\hat{x}$ side and a Au nucleus traveling along the $-\hat{z}$ direction on the $-\hat{x}$ side collides at $z=0$.  Panel (a): configuration before collision. Panel (b): overlap between Cu and Au in the transverse ($x$-$y$) plane at $z=0$, with arrows denoting the initial pressure gradient of the QGP. Panel (c): initial geometry of the QGP in the reaction ($x$-$z$) plane at $y=0$, with arrows denoting the longitudinal flow velocity field of the QGP.}
\label{fig1:sketch of cuau}
\end{figure}

As illustrated in Fig.~\ref{fig1:sketch of cuau}, the directed flow in Cu+Au collisions can originate from various sources. As shown in panel (b), the asymmetric energy deposition from Au and Cu nuclei in the transverse plane would cause different pressure gradients between $+\hat{x}$ and $-\hat{x}$ directions. Meanwhile, as shown in panel (c), the interactions between the spectator nucleons and the QGP can generate a tilt of the QGP in the reaction plane, which further leads to an asymmetry in transverse planes at forward and backward rapidity regions and thus finite $v_1$. Additionally, the longitudinal velocity field can transport matter between different rapidities and modify the asymmetries generated by the previous two sources. Since Cu and Au nuclei also deposit different amount of momentum inside the QGP between $+\hat{z}$ and $-\hat{z}$ directions, one expects to see different magnitudes of $v_1$ between forward and backward rapidities. In this work, we will extend the well established TRENTo model~\cite{Moreland:2014oya} from 2-dimensional (D) to 3D space, incorporating all these effects in the initial condition of the QGP. By combining this initial condition with the (3+1)-D viscous hydrodynamic model CLVisc~\cite{Pang:2018zzo,Wu:2021fjf}, we are able to provide a good description of the pseudorapidity, transverse momentum, and centrality dependences of the hadron $v_1$ in Cu+Au collisions at a center-of-mass nucleon-nucleon collision energy of $\snn=200$~GeV. We also provide predictions for the elliptic flow coefficient ($v_2$) of hadrons and the global polarization of hyperons in these asymmetric collisions. We find that effects of the tilted QGP geometry on these flow and polarization observables appear stronger in Cu+Au than in symmetric Au+Au collisions, and different observables show different sensitivities to the longitudinal flow velocity gradient of the QGP. Therefore, a combined study of these observables in Cu+Au collisions provides a unique opportunity for constraining the initial condition of high-energy nuclear collisions.

The rest of this article is organized as follows. In Sec.~\ref{section2}, we discuss the theoretical framework we develop for studying collective flow coefficients and global polarization in heavy-ion collisions, including an extension of the TRENTo initial condition to the 3D space, and a hydrodynamic simulation of the QGP evolution. In Sec~\ref{section3}, we present the numerical results on the directed flow and elliptic flow coefficients of hadrons, and the global polarization of $\Lambda$ and $\bar{\Lambda}$ hyperons, and investigate their dependences on different components of our initial condition model. In the end, we summarize in Sec.~\ref{section4}.

%%%%%%% section-2

\section{Initialization and evolution of the QGP}
\label{section2}

In this section, we will first introduce how we extend the 2D TRENTo model to a 3D space for initializing the energy and net baryon density of the QGP. Then, we will discuss the evolution of the QGP using the (3+1)-D viscous hydrodynamic model CLVisc and its particlization using the Cooper-Frye formalism. Model parameters used in this work will be listed.

\subsection{Modified TRENTo initial condition with tilted geometry}
\label{section2_sub1}

%%%  Fig-1 
\begin{figure*}[tbp!]
\begin{center}
\includegraphics[width=0.24\linewidth]{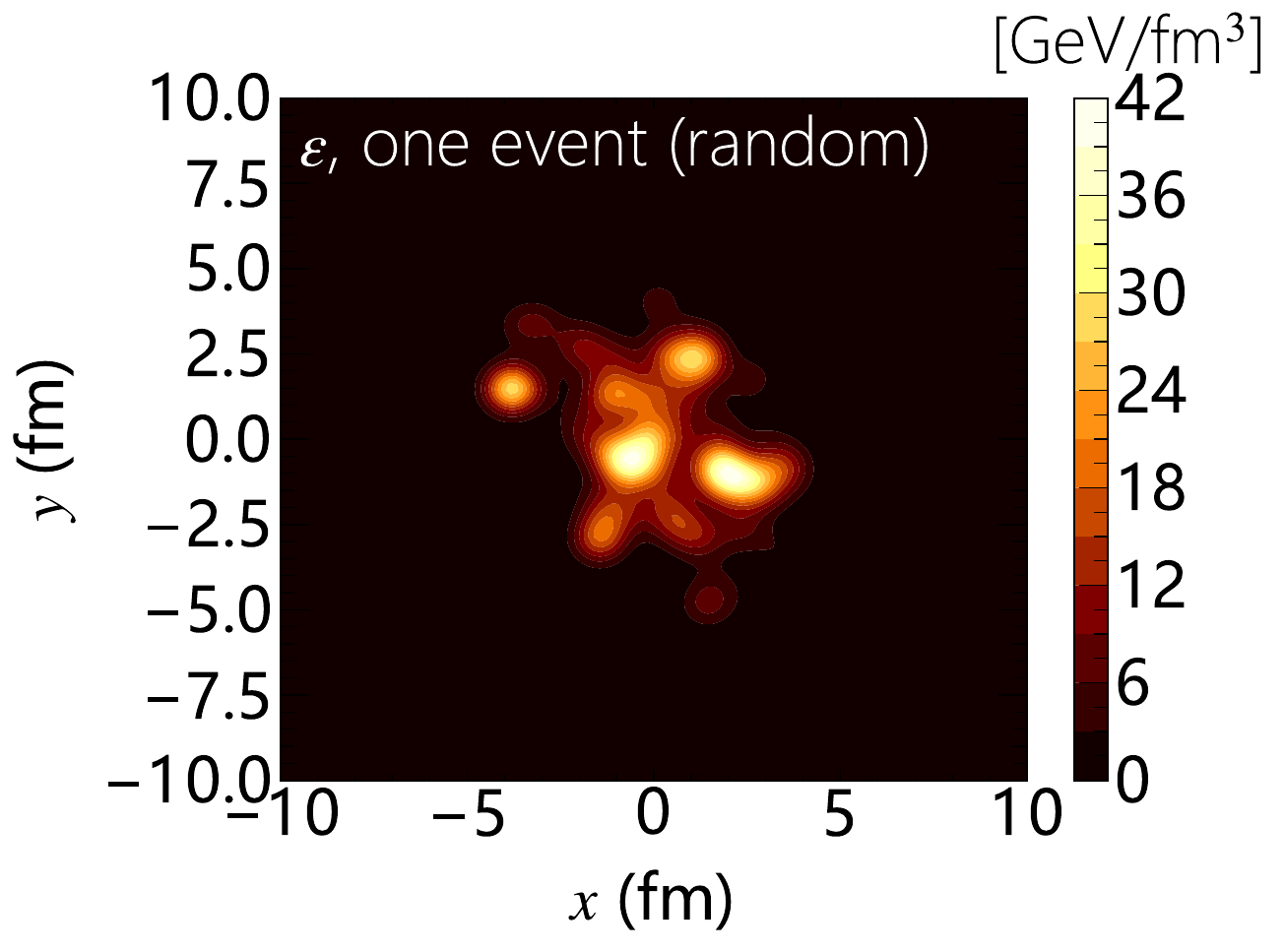}~
\includegraphics[width=0.24\linewidth]{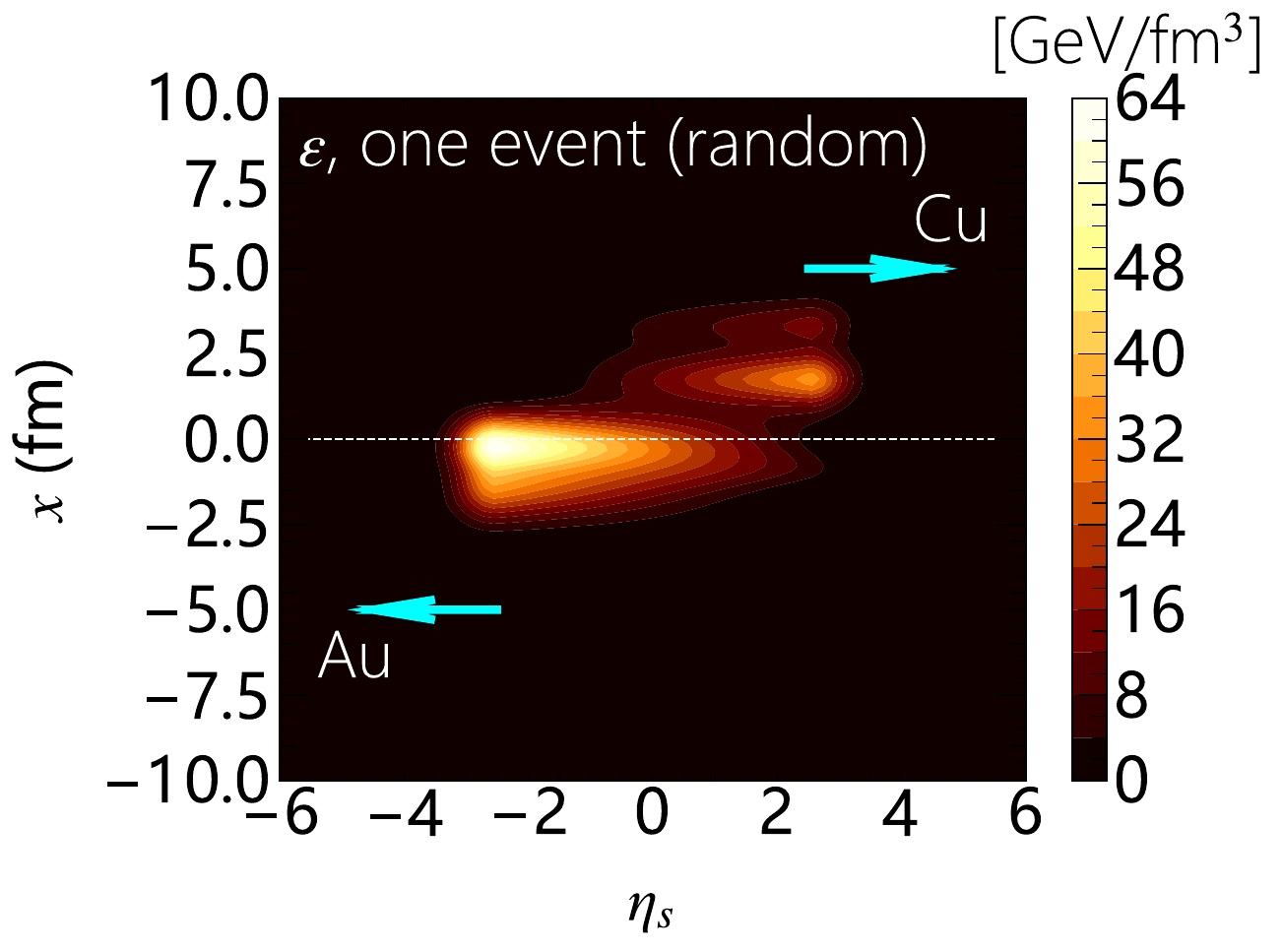}~
\includegraphics[width=0.24\linewidth]{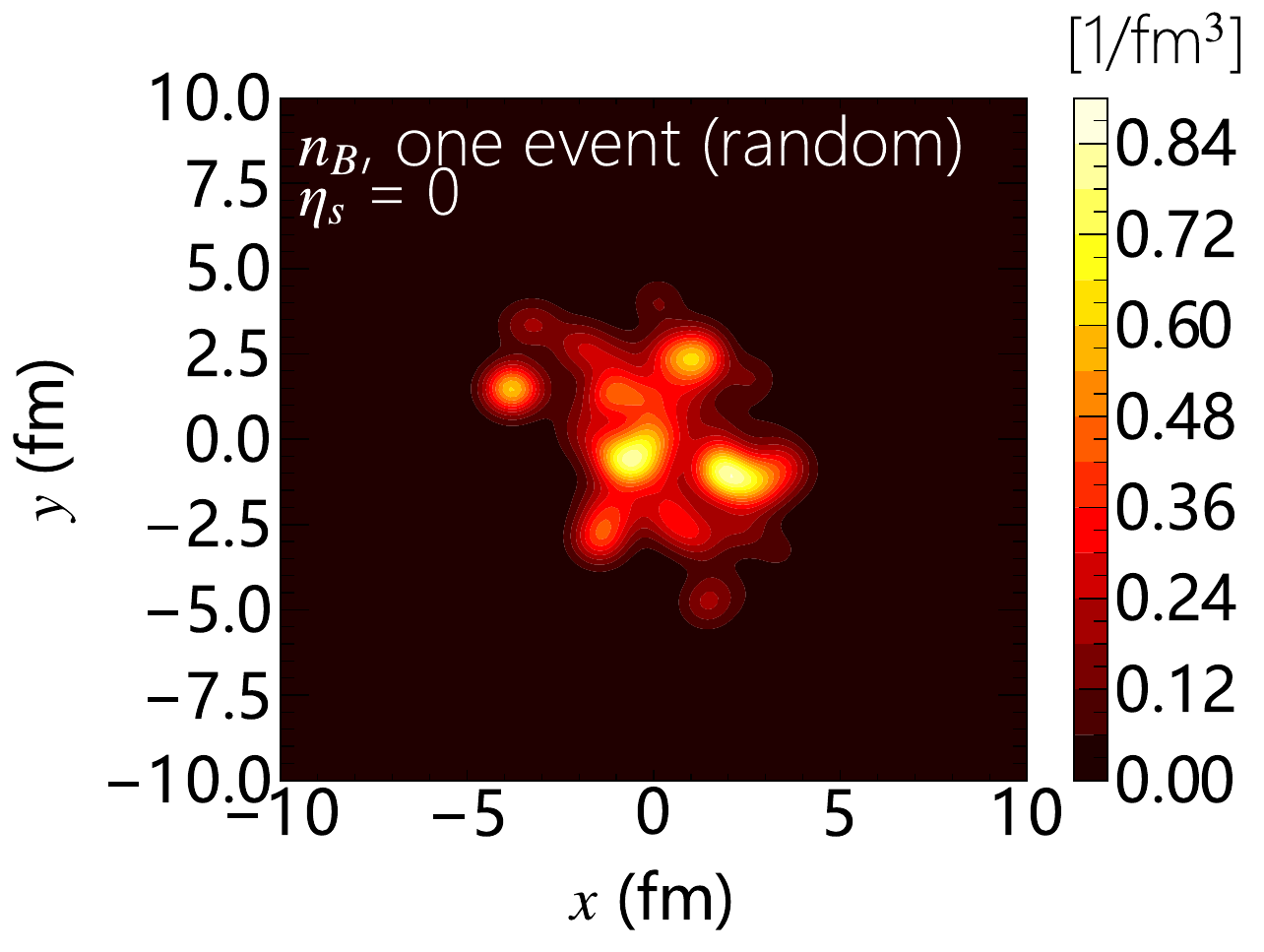}~
\includegraphics[width=0.24\linewidth]{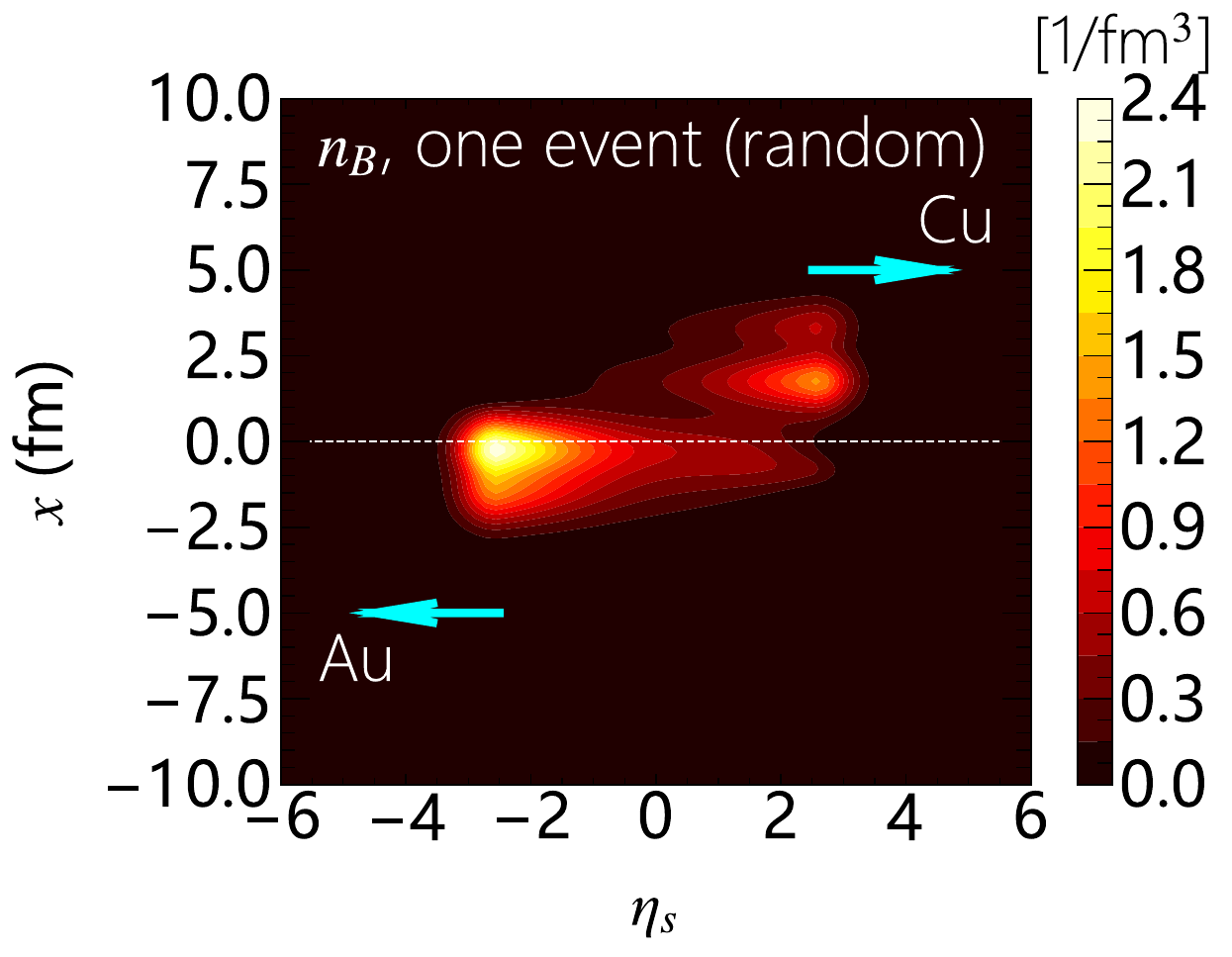} \\
\includegraphics[width=0.24\linewidth]{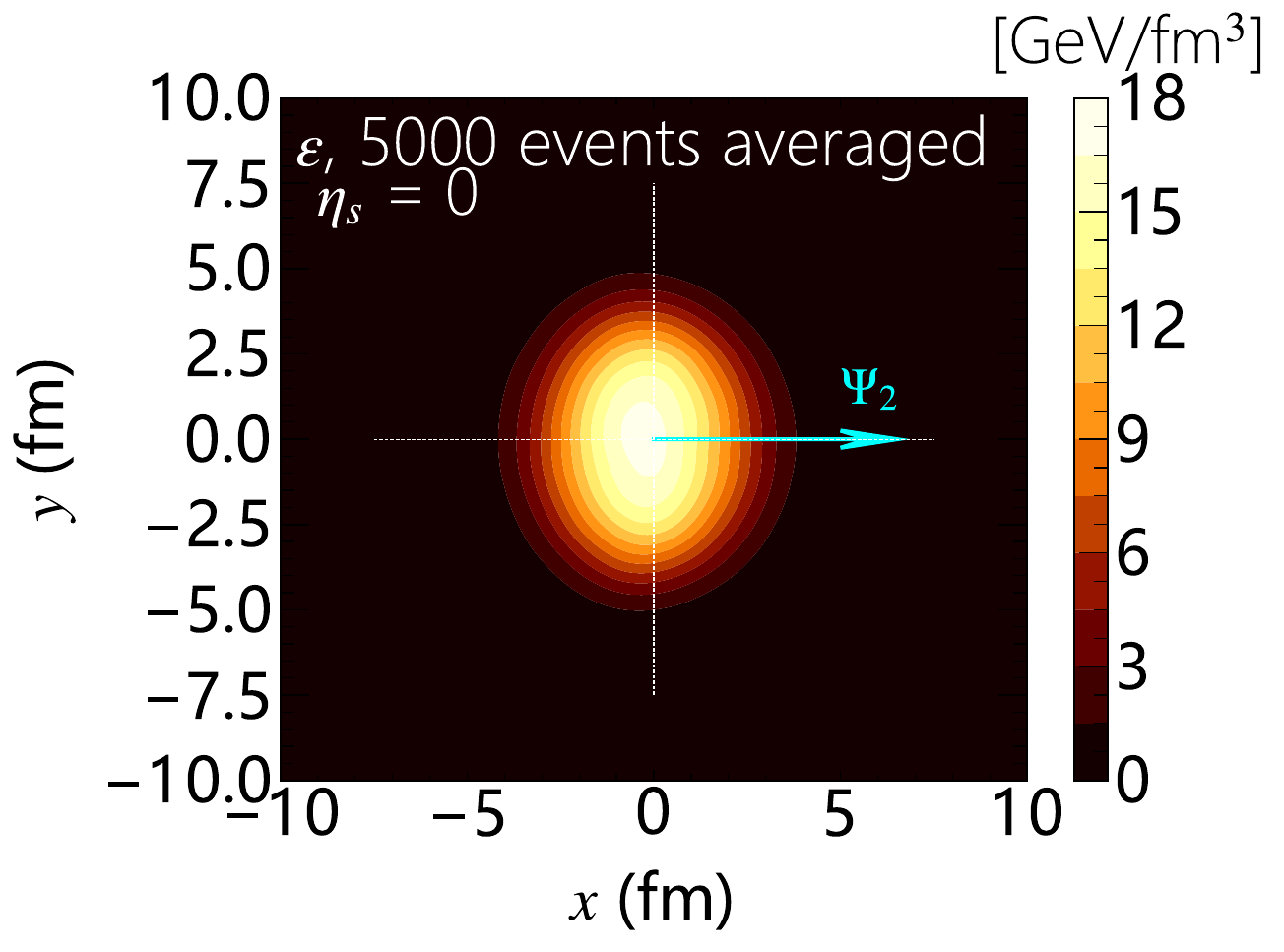}~
\includegraphics[width=0.24\linewidth]{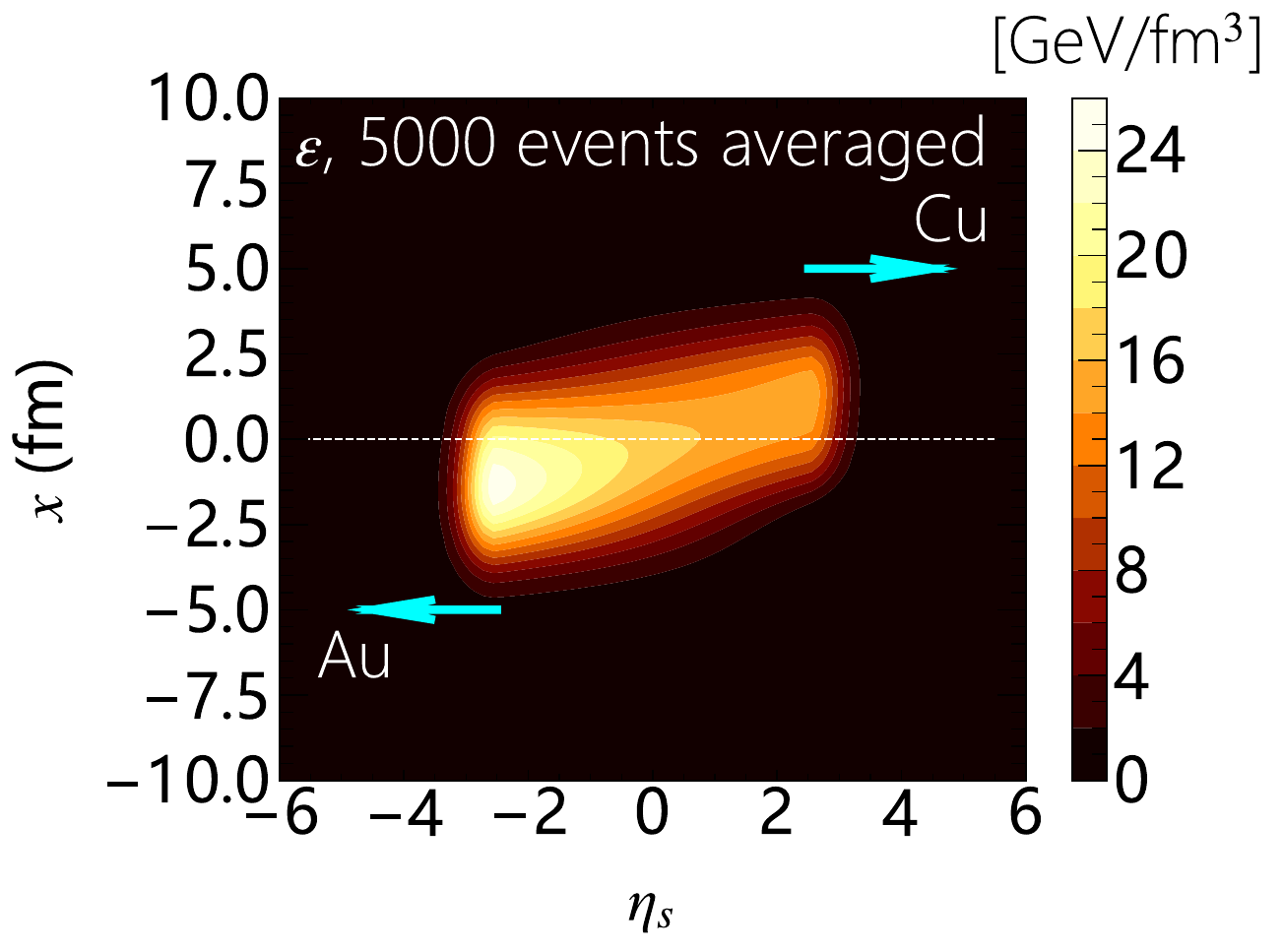}~
\includegraphics[width=0.24\linewidth]{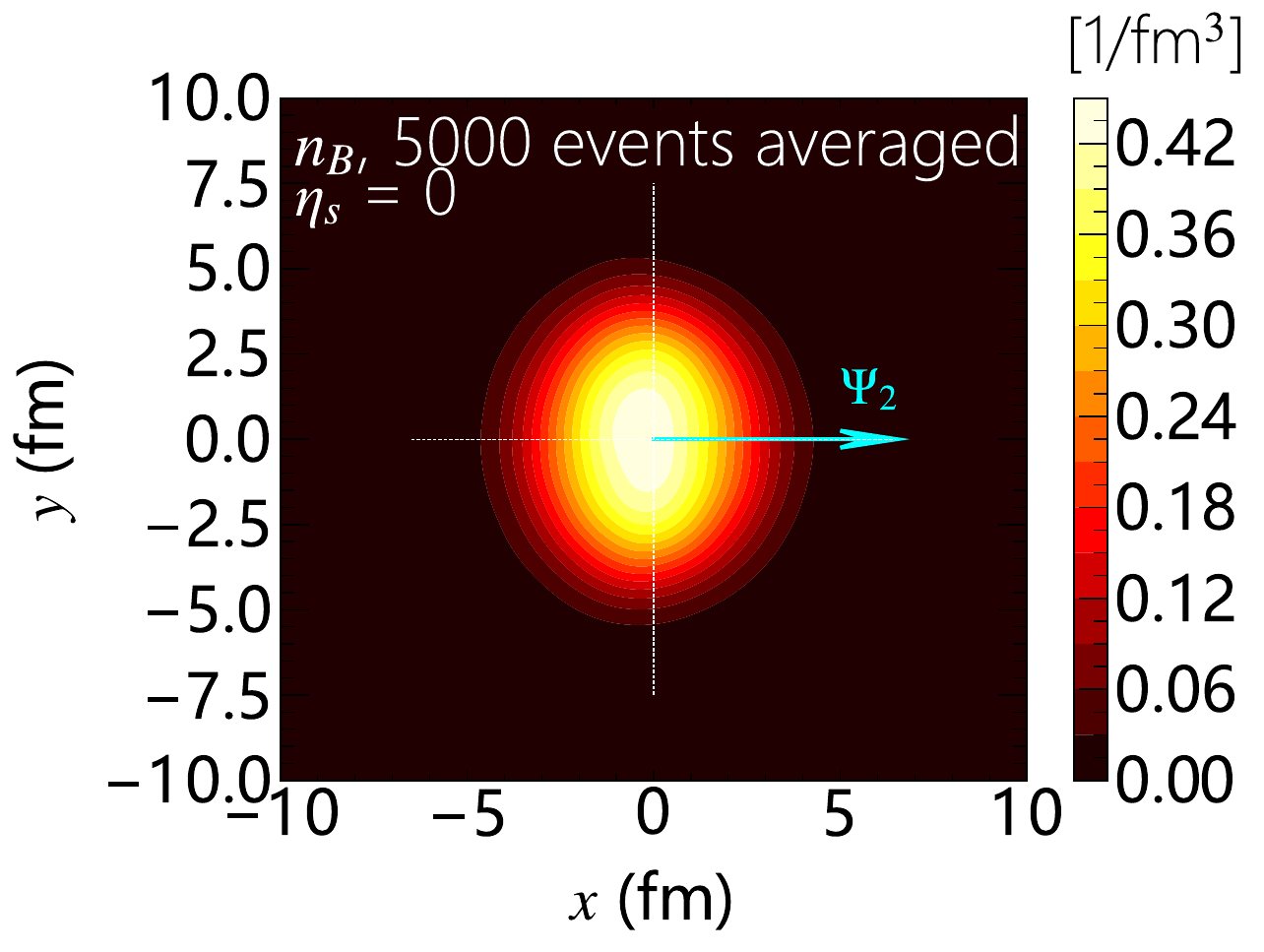}~
\includegraphics[width=0.24\linewidth]{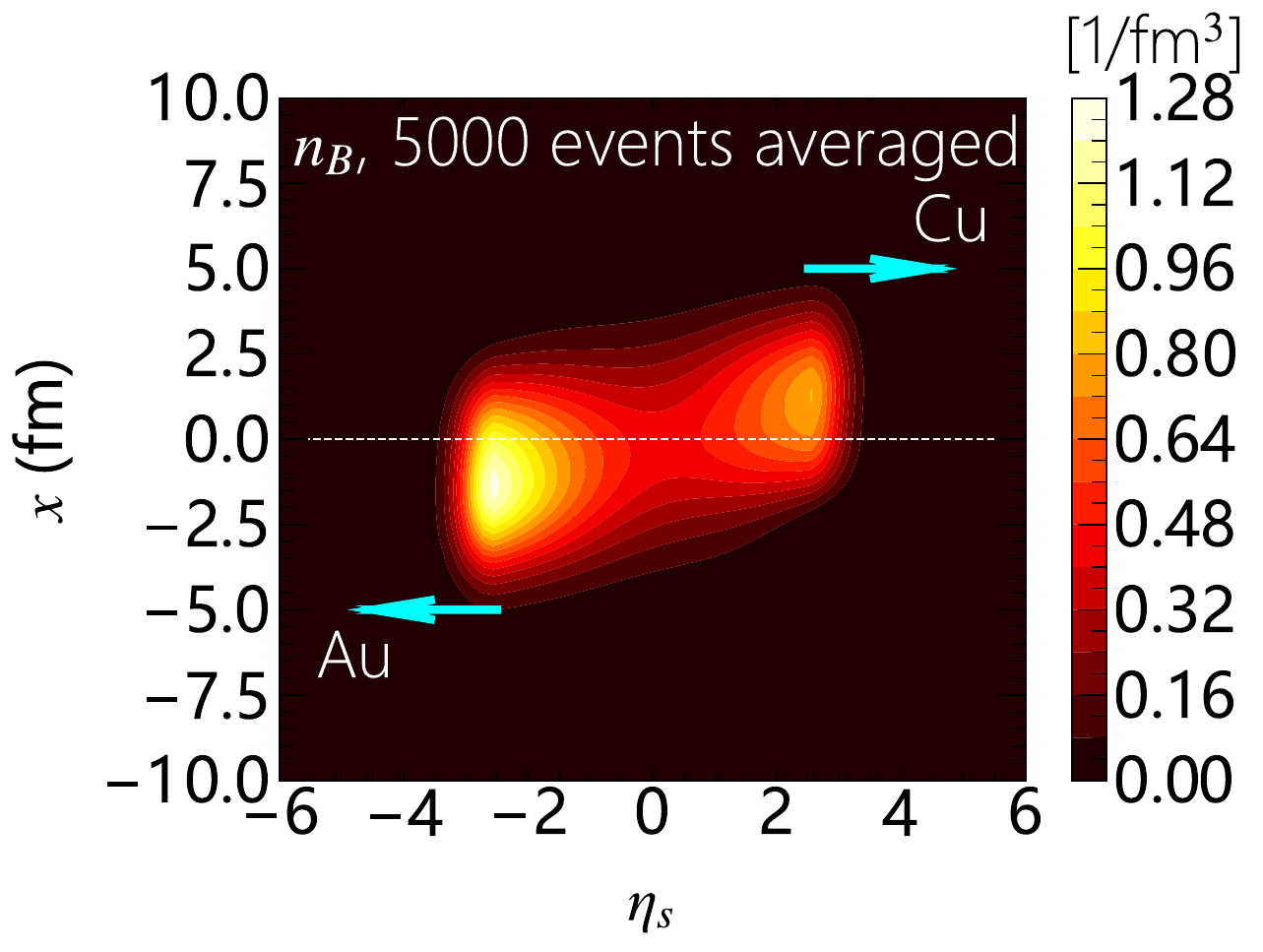}
\end{center}
\caption{(Color online) 
The initial energy density (left four panels) and net baryon density (right four panels) distributions in 10-40\% Cu+Au collisions at $\sqrt{s_\text{NN}}=200$~GeV, upper panels for one random event, and lower panels for smooth profiles averaged over 5000 events. The first and third columns are displayed on the $x$-$y$ plane with $\eta_\text{s}=0$, while the second and fourth columns are displayed on the $\eta_\text{s}$-$x$ plane with $y=0$.}
\label{f:ed}
\end{figure*}

%A refined 3D TRENTo model is performed to generate the initial conditions for the hydrodynamic evolution of the QGP, featuring a counter-clockwise tilt in the reaction plane with respect to the beam (longitudinal) direction and an non-zero longitudinal velocity gradient.

The 2D TRENTo model~\cite{Moreland:2014oya,Bernhard:2016tnd} is a successful tool for generating the initial transverse profile the QGP in relativistic heavy-ion collisions. We start with the participant thickness function of a nucleus $A$ generated in TRENTo as:
\begin{equation}
\label{eq:thickness}
T_A(x,y)=\sum_{i=1}^{N_\mathrm{part}}\gamma_i T_p(x-x_i,y-y_i),
\end{equation}
where $(x,y)$ denote the transverse coordinates, $N_\mathrm{part}$ is the number of participant nucleons in $A$, $T_p$ represents the nucleon thickness function, parametrized using a normalized 2D Gaussian distribution with a width parameter $w$, and $\gamma_i$ is sampled from a gamma distribution with unit mean and variance $1/k$ for participant nucleon $i$ to account for minimum-bias proton-proton multiplicity fluctuations. In this work, we use the default values $w=0.5$~fm and $k=1$ in the TRENTo package~\cite{trento:2d}.

The nucleon density inside a nucleus is given by the Woods-Saxon distribution:
\begin{equation}
\begin{aligned}
\rho(r)=\frac{\rho_{0}}{1+\exp\left[(r-R_{0})/d_{0}\right]},
\label{eq:WS}
\end{aligned}
\end{equation}
where $\rho_0$ is the average nucleon density, $r=\sqrt{x^{2}+y^{2}+z^{2}}$ is the radial position with $z$ denoting the longitudinal coordinate along the beam direction, $R_{0}$ is the radius of the nucleus and $d_{0}$ is the surface diffusiveness parameter~\cite{Alver:2008aq,Loizides:2017ack}. For collisions between two nuclei with an impact parameter $b$, the nucleon positions inside the projectile (target) nuclei are sampled according to Eq.~(\ref{eq:WS}) with a $b/2$ ($-b/2$) shift. Two nucleons, one from the projectile and one from the target, collide if the probability 
\begin{equation}
P_\mathrm{coll}(b)=1-\exp[-\sigma_{gg}T_{pp}(b)]
\end{equation}
is satisfied, where
\begin{equation}
T_{pp}(b)=\int dx dy T_p(x,y) T_p(x-b,y),
\end{equation}
and 
$\sigma_{gg}$ is fixed by reproducing the inelastic nucleon-nucleon cross section 
\begin{equation}
\sigma_\mathrm{NN}^\mathrm{inel}=\int db\, 2\pi b\, P_\mathrm{coll}(b),
\end{equation}
e.g., 42~mb for $\sqrt{s_\mathrm{NN}}=200$~GeV. The thickness functions of the participant nucleons are then substituted into Eq.~(\ref{eq:thickness}) for the participant thickness function of a nucleus. The parameters of the Woods-Saxon distributions of Cu and Au nuclei are listed in Table~\ref{t:parameters}.

\begin{table}[h]
\begin{center}
\caption{Parameters of the Woods-Saxon distributions for Cu and Au nuclei~\cite{Loizides:2017ack}. }
\begin{tabular}{c c c c c}
\hline\hline
nucleus     & $A$          & $\rho_{0}$ [1/fm$^{3}$]      & $R_{0}$~[fm]    & $d_{0}$ [fm]    \\ \hline
Cu          & 63         & 0.17                      & 4.20      & 0.546       \\
Au          & 197        & 0.17                      & 6.38      & 0.546       \\
\hline\hline
\label{t:parameters}
\end{tabular}
\end{center}
\end{table}

Using the participant thickness functions provided by TRENTo, $T_1$ for projectile nucleus and $T_2$ for target nucleus, we construct the initial transverse entropy density distributions at different spacetime rapidities ($\eta_\text{s}$) as
\begin{equation}
\begin{aligned}
s_0(x, y, \eta_\text{s}) = K \tilde{s}_{0}(x,y,\eta_\text{s}) H(\eta_\text{s}), 
\label{eq:entropy}
\end{aligned}
\end{equation}
with
\begin{equation}
\begin{aligned}
\tilde{s}_{0}(x,y,\eta_\text{s})=&~\left[\frac{T^{p}_{1}(x,y)+T^{p}_{2}(x,y)}{2}\right]^{\frac{1}{p}} \\
+&~{H_\text{t}}\left[\frac{T^{p}_{1}(x,y)-T^{p}_{2}(x,y)}{2}\right]^{\frac{1}{p}}\tan\left(\frac{\eta_\text{s}}{\eta_\text{t}}\right),
\label{eq:mnccnu}
\end{aligned}
\end{equation}
and
\begin{equation}
\begin{aligned}
H(\eta_\text{s})=\exp\left[-\frac{(|\eta_\text{s}|-\eta_\text{w})^{2}}{2\sigma^{2}_{\eta}}\theta(|\eta_\text{s}|-\eta_\text{w}) \right].
\label{eq:Hs}
\end{aligned}
\end{equation}
In Eq.~(\ref{eq:entropy}), $K$ is an overall normalization factor which will be adjusted to describe the $p_{\textrm{T}}$ spectra of identified particles ($\pi^{+}$, $K^{+}$, $p$, and $\bar{p}$) in the most central Cu+Au collisions. The first part of Eq.~(\ref{eq:mnccnu}) is taken from the 2D TRENTo model, while the second part is introduced from our earlier work~\cite{Jiang:2021foj,Jiang:2021ajc,Jiang:2023vxp} to account for the imbalanced entropy deposition between the projectile and target, in which $H_\text{t}$ reflects the overall strength of imbalance and the function $\tan(\eta_\text{s}/\eta_\text{t})$ determines how the imbalance varies with $\eta_\text{s}$. This parametrization generates a counterclockwise tilt of the QGP fireball in the reaction plane with respect to the beam direction, which can result from the interactions between spectators and the QGP, and is preferred by the phenomenology of directed flow in heavy-ion collisions~\cite{Bozek:2011ua,Bozek:2012hy,Bozek:2016kpf}. In this work, we keep using a constant parameter $\eta_\text{t}=8.0$ which provides a good description of the charged particle $v_1$ in Au+Au collisions at RHIC~\cite{Jiang:2021foj,Jiang:2023fad,Jiang:2023vxp}, and apply $H_\text{t}=0.8b$/fm for describing the centrality dependence of $v_1$ in Cu+Au collisions later. The parameter $p$ in Eq.~(\ref{eq:mnccnu}) is taken as 1 in the present study, corresponding to the wounded nucleon model. Other choices, as proposed in the TRENTo work~\cite{Moreland:2014oya,Bernhard:2016tnd}, will be explored in our future efforts. The envelop function $H$, given by Eq.~(\ref{eq:Hs}), is introduced to describe the plateau structure in the longitudinal distribution of emitted hadrons, in which  $\eta_\text{w}$ controls the width of the central rapidity plateau and $\sigma_{\eta}$ determines the width (speed) of the Gaussian decay outside the plateau region~\cite{Pang:2018zzo}.

Similarly, we construct the net baryon density as~\cite{Denicol:2018wdp,Bozek:2022svy} 
\begin{align}
\label{eq:nb} n_{B}(x,y,\eta_\text{s})  = \frac{1}{N_{B}} \tilde{s}_{0}(x,y,\eta_\text{s}) H(\eta_\text{s})  H_{B}(\eta_\text{s}) ,
\end{align}
with $N_{B}$ a factor that normalizes the integrated $n_{B}$ to the number of participant nucleons $N_{\textrm{part}}$. To model the accumulation of baryons in the forward and backward rapidity regions, we incorporate the following distribution function of baryon density along the longitudinal direction~\cite{Bozek:2022svy,Jiang:2023fad},
\begin{equation}
\begin{aligned}
H_{B}(\eta_\text{s})=\exp\left[-\frac{(\eta_\text{s}-\eta_{n})^{2}}{2\sigma^{2}_{n}}\right]+\exp\left[-\frac{(\eta_\text{s}+\eta_{n})^{2}}{2\sigma^{2}_{n}}\right],
\label{eq:heta}
\end{aligned}
\end{equation}
where parameters $\eta_{n}$ and $\sigma_{n}$ will be calibrated by the $p_\text{T}$ spectra of protons and antiprotons. 

Since the directed flow is sensitive to the fluid velocity field in the longitudinal direction~\cite{Shen:2020jwv,Jiang:2023fad}, we need to extend the initialization model beyond the Bjorken approximation for the fluid velocity. 
The initial energy-momentum tensor components related to the longitudinal direction are given by~\cite{Shen:2020jwv, Ryu:2021lnx, Alzhrani:2022dpi}
\begin{align}
\label{eq:Ttautau}
T^{\tau\tau}&=\varepsilon_{0}(x,y,\eta_\text{s})\cosh(y_\text{L}) \, ,\\
\label{eq:Ttaueta}
T^{\tau\eta_\text{s}}&=\frac{1}{\tau_{0}}\varepsilon_{0}(x,y,\eta_\text{s})\sinh(y_\text{L})  \, ,
\end{align}
where $\varepsilon_{0}$ is the initial energy density converted from the entropy density $s_{0}$ via the equation of state. The rapidity variable is modeled as $y_\text{L} = f_{v} y_{\textrm{loc}}^{\text{cm}}$, where
\begin{equation}
\begin{aligned}
y_{\textrm{loc}}^{\text{cm}}=\textrm{arctanh} \left[\frac{T_{1}-T_{2}}{T_{1}+T_{2}} \tanh (y_{\textrm{beam}})\right]
\label{eq:ycm}
\end{aligned}
\end{equation}
is the local center-of-mass rapidity of participants from projectile and target nuclei. In the above equations, $y_{\textrm{beam}} = \textrm{arccosh}[\sqrt{s_{\textrm{NN}}}/(2m_{\textrm{N}})]$ is the rapidity of the colliding beam, with $m_{\textrm{N}}$ the nucleon mass, and the parameter $f_{v}\in[0,1]$ controls the fraction of the local center-of-mass rapidity deposited from the participant nucleons into the QGP medium. Therefore, $f_{v}$ determines the magnitude of the longitudinal flow velocity gradient. The Bjorken flow scenario with $y_\text{L}=0$~\cite{Shen:2020jwv} can be recovered when $f_{v}=0$ is taken. With Eqs.~(\ref{eq:Ttautau}) and~(\ref{eq:Ttaueta}), the initial fluid velocity in the $\eta_\text{s}$ direction is given by $v_{\eta_\text{s}}=T^{\tau\eta_\text{s}}/(T^{\tau\tau}+P)$, in which $P$ denotes pressure.
In the present work, the initial fluid velocity in the transverse plane is assumed to be zero by setting $T^{\tau x} = T^{\tau y} = 0$. Using the modified energy-momentum tensor components above, we recalculate the initial local energy density of the QGP beyond the Bjorken approximation.

\begin{table}[h]
\centering
\vline
\caption{Parameters of the initial conditions for Cu+Au collisions at $\sqrt{s_\mathrm{NN}}=200$~GeV.}
\begin{tabular}{ c c c c c c c c c c }
\hline\hline
System & ~$K$~   & $\tau_0$ [fm] & $\sigma_\eta$ & $\eta_\text{w}$ & $\sigma_{n}$ &$\eta_{n}$ & ~$N_{B}$~ & $f_v$  \\ \hline
Cu+Au  & ~46~   & 0.6 & 0.4 & 2.5   & 2.0  &3.4 &~20~ &0.05  \\ \hline\hline
\end{tabular}
\label{table:parameters}
\end{table}

In Tab.~\ref{table:parameters}, we list the parameters we use in this work to initialize the QGP medium. The first six parameters ($K$, $\tau_0$, $\sigma_\eta$, $\eta_\text{w}$, $\eta_{n}$ and $\sigma_{n}$) 
are adjusted based on the $p_\text{T}$ spectra of the charged particle yields in the most central collisions. The last parameter ($f_{v}$) is adjusted according to the directed flow coefficients of charged particles. 

Using these parameters, we present in Fig.~\ref{f:ed} the initial energy density and net baryon density on the $\eta_\text{s}$-$x$ and $x$-$y$ planes at $\tau_{0}=0.6$~fm for 10-40\% Cu+Au collisions at $\snn=200$~GeV. The upper panel shows results for one random event, and the lower panel shows results averaged over 5000 events. For the lower panel, the average is performed after the second order participant plane angle $\Psi_{2}$ of each event is rotated to the $x$-axis. From the second and the fourth column of Fig.~\ref{f:ed}, one can clearly see a counterclockwise tilt of the QGP fireball in the participant plane relative to the longitudinal direction from our initialization scheme. Stronger tilt is expected in more peripheral collisions due to stronger impact from spectators on the QGP. Compared to the energy density profile, the net baryon distribution exhibits a stronger shift towards larger rapidity regions. This is in line with the string model of the initial state: while baryon density deposition is driven by valence quarks, energy density deposition originates from string melting that involves both valence and sea quarks. In addition, we see stronger energy and baryon depositions towards the backward direction than the forward direction due to larger total energy and baryon number carried by the Au nuclei than Cu nuclei in these asymmetric collisions. From the lower panels of the first and third columns, one can also observe stronger energy and net baryon deposition in the $-\hat{x}$ direction than in the $+\hat{x}$ direction in the $x$-$y$ at mid-pseudorapidity in these asymmetric Cu+Au collisions. The asymmetry in the energy density distribution here will affect the directed flow of soft hadrons, and the asymmetry in the net baryon density distribution will affect the baryon and anti-baryon productions at different locations inside the QGP~\cite{Bozek:2022svy}. In this study, we use the smooth initial condition, as shown in the lower panels here, for the subsequent calculations. Effects of the initial-state fluctuations on the final-state hadron $v_1$ will be explored in our future work.  

\begin{figure}[tbp!]
\begin{center}
\includegraphics[width=0.85\linewidth]{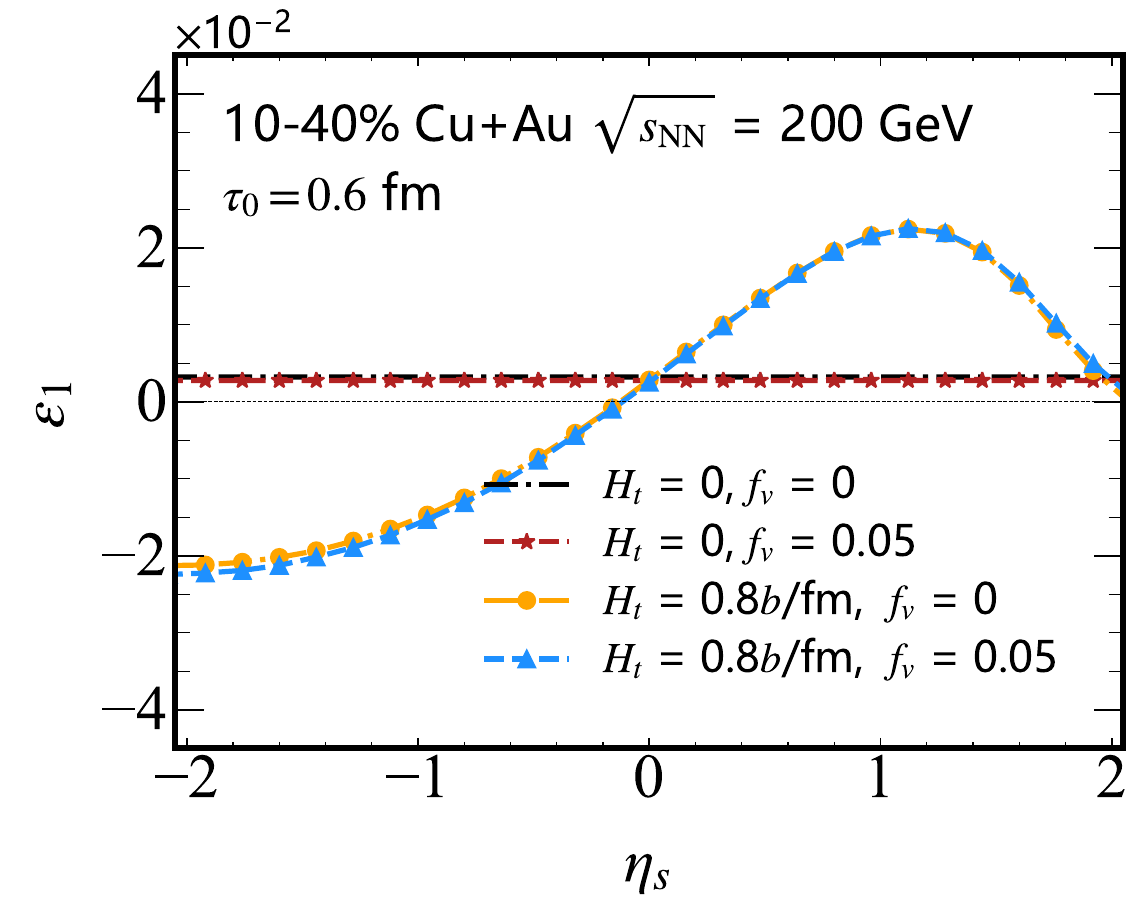} \\
\end{center}
\caption{(Color online) The first order eccentricity of the smooth energy density distribution as a function of spacetime rapidity for 10-40\% Cu+Au collisions at $\sqrt{s_\mathrm{NN}}=200$~GeV, compared between different parameter setups of our initial condition model.}
\label{f:cuau200ecc1}
\end{figure}

In Fig.~\ref{f:cuau200ecc1}, we present the first order eccentricity coefficient, $\varepsilon_1$, as a function of spacetime rapidity, utilizing the smooth initial condition. Results from different parameter settings are compared. This helps illustrate how our model setups affect the asymmetric distribution of the energy density of the QGP. The first order eccentricity vector is defined as~\cite{Qiu:2011iv,Shen:2020jwv}
\begin{equation}
\begin{aligned}
\vec{\mathcal{E}}_{1}\equiv\varepsilon_{1}(\eta_\text{s})e^{i\Psi_{1}(\eta_\text{s})}=
- \frac{\int d^{2}r \widetilde{r}^{3} e^{i\widetilde{\phi}}\varepsilon(r,\phi,\eta_\text{s})}{\int d^{2}r \widetilde{r}^{3} \varepsilon(r,\phi,\eta_\text{s})},
\label{eq:ecc1}
\end{aligned}
\end{equation}
in which the transverse radius $\tilde{r}$ and azimuthal angle $\tilde{\phi}$ are measured with respect to the center-of-mass $(x_{0}, y_{0})$ of a slice of the medium at a given spacetime rapidity,
\begin{equation}
\begin{aligned}
x_{0}(\eta_\text{s}) = \frac{\int d^{2} r x \varepsilon(r,\phi,\eta_\text{s})}{\int d^{2}r \varepsilon(r,\phi,\eta_\text{s})},
\label{eq:xy1}
\end{aligned}
\end{equation}
\begin{equation}
\begin{aligned}
y_{0}(\eta_\text{s}) = \frac{\int d^{2} r y \varepsilon(r,\phi,\eta_\text{s})}{\int d^{2}r \varepsilon(r,\phi,\eta_\text{s})}.
\end{aligned}
\end{equation}
Then, we have $\widetilde{r}=\sqrt{(x-x_{0})^{2}+(y-y_{0})^{2}}$ and $\widetilde{\phi}=\arctan[(y-y_{0})/(x-x_{0})]$. In the end, $\varepsilon_1$ in Eq.~(\ref{eq:ecc1}) gives the magnitude of the first order eccentricity, while $\Psi_1$ gives the corresponding participant plane angle. 
Compared to Au+Au collisions~\cite{Jiang:2021ajc}, the energy density deposited by non-central Cu+Au collisions in the transverse plane shows clear asymmetry between $+\hat{x}$ and $-\hat{x}$ directions. When the tilt geometry is not introduced, i.e., $H_\mathrm{t}=0$, a positive $\varepsilon_1$ is seen across the entire $\eta_\text{s}$ region, implying a wider distribution of energy density deposited on the Au ($-\hat{x}$) side. This would result in a larger pressure gradient on the Cu ($+\hat{x}$) side and therefore a positive value of $v_1$ in the momentum space later. 
Note that such non-zero value of $\varepsilon_1$ can appear in Au+Au collisions only when event-by-event fluctuations are taken into account. The titled geometry (nonzero $H_\text{t}$) of the QGP introduces additional asymmetry in the transverse plane, which depends on spacetime rapidity. For symmetric Au+Au collisions, the value of $\varepsilon_1$ would be negative at $\eta_\text{s}<0$ while positive at $\eta_\text{s}>0$, crossing zero at $\eta_\text{s}=0$~\cite{Jiang:2021ajc}. However, due to the asymmetric energy deposition from Au and Cu nuclei as discussed above for the $H_{t}=0$ case, the $\varepsilon_1(\eta_\text{s})$ function of Au+Au collisions is shifted towards the $+\hat{x}$ direction for Cu+Au collisions, and therefore crosses zero at negative $\eta_\text{s}$.
The asymmetry of the initial energy density distribution shown here is an essential, though not the only, source of the final state directed flow coefficient discussed in the next section. The $f_v$ parameter, which governs the longitudinal flow velocity of the QGP, has negligible impact on $\varepsilon_1$ here, but will be crucial for $v_1$ later. The net baryon density distribution will also affect the $v_1$ of baryons.

\subsection{Hydrodynamic evolution}
\label{section2_sub2}

Starting with the initial condition constructed in the previous subsection, we utilize a (3+1)-D viscous hydrodynamic model CLVisc~\cite{Pang:2018zzo,Wu:2021fjf} to describe the further evolution of the QGP medium and its particlization. 
Details of the CLVisc model can be found in Refs.~\cite{Pang:2018zzo,Wu:2021fjf}. Here, we briefly summarize its main functions applied in this work. 

The hydrodynamic evolution is based on the following conservation equations: 
\begin{align}
\nabla_{\mu} T^{\mu\nu}&=0 \, ,\\
\nabla_{\mu} J^{\mu}&=0  \, ,
\end{align}
where the energy-momentum tensor $T^{\mu\nu}$ and the net baryon current $J^{\mu}$ are defined as
\begin{align}
T^{\mu\nu} &= \varepsilon u^{\mu}u^{\nu} - P\Delta^{\mu\nu} + \pi^{\mu\nu}\,, \\	
J^{\mu} &= nu^{\mu}+V^{\mu}\,,
\end{align}
with $\varepsilon$, $P$, $u^{\mu}$, $\pi^{\mu\nu}$, $n$, $V^{\mu}$ the local energy density, pressure, flow velocity field, shear stress tensor, net baryon density, and baryon diffusion current, respectively.
The projection tensor is given by $\Delta^{\mu\nu} = g^{\mu\nu}-u^{\mu}u^{\nu}$ with the metric tensor $g^{\mu\nu} = \text{diag} (1,-1,-1,-1)$. The dissipative currents $\pi^{\mu\nu}$ and $V^{\mu}$ are given by the Israel-Stewart-like second order hydrodynamic expansion~\citep{Denicol:2018wdp}, but effects of bulk viscosity is not included in the present study yet.
Two key quantities -- specific shear viscosity ($C_{\eta_\text{v}}$) and baryon diffusion coefficient ($\kappa_B$) embedded in $\pi^{\mu\nu}$ and $V^{\mu}$ -- are related to shear viscosity ($\eta_\text{v}$) and parameter $C_B$ via
\begin{align}
C_{\eta_\text{v}} &= \frac{\eta_\text{v} T}{\varepsilon+P}, \label{eq:C_shear}\\
\kappa_B &= \frac{C_B}{T}n\left[\frac{1}{3} \cot \left(\frac{\mu_B}{T}\right)-\frac{nT}{\varepsilon+P}\right] \,, \label{eq:CB}
\end{align}
in which $T$ is the local temperature and $\mu_B$ is the baryon chemical potential. They are related to relaxation times via  $\tau_{\pi} = 5C_{\eta_\text{v}}/T$ and $\tau_V = C_B/T$.
In this work, we set $C_{\eta_\text{v}}=0.08$ and ${C_B}=0.4$.

The NEOS-BQS equation of state (EOS) is adopted for solving the hydrodynamic equations ~\cite{Monnai:2019hkn,Monnai:2021kgu}, which provides a smooth crossover between the QGP and the hadron phase under the conditions of strangeness neutrality ($n_S=0$) and electric charge density $n_Q = 0.4n_B$. The freezeout hypersurface is determined by the energy density at $\varepsilon_{\text{frz}}$= 0.4~GeV/fm$^3$, on which the spectra of different hadron species are determined based on the Cooper-Frye formalism. The resonance decay contribution is included in evaluating the collective flows of identified particles. Effects of hadronic scatterings after the QGP phase have not been included in the present work. Detailed discussions on the hadronization processes can be found in Refs.~\cite{Wu:2021fjf,Jiang:2023fad}.

%%%%%%  Section 3 
\section{Numerical results}
\label{section3}

\subsection{Transverse momentum spectra and multiplicity distributions}
\label{section3_sub1}

\begin{figure}[tbp!]
\begin{center}
\includegraphics[width=0.9\linewidth]{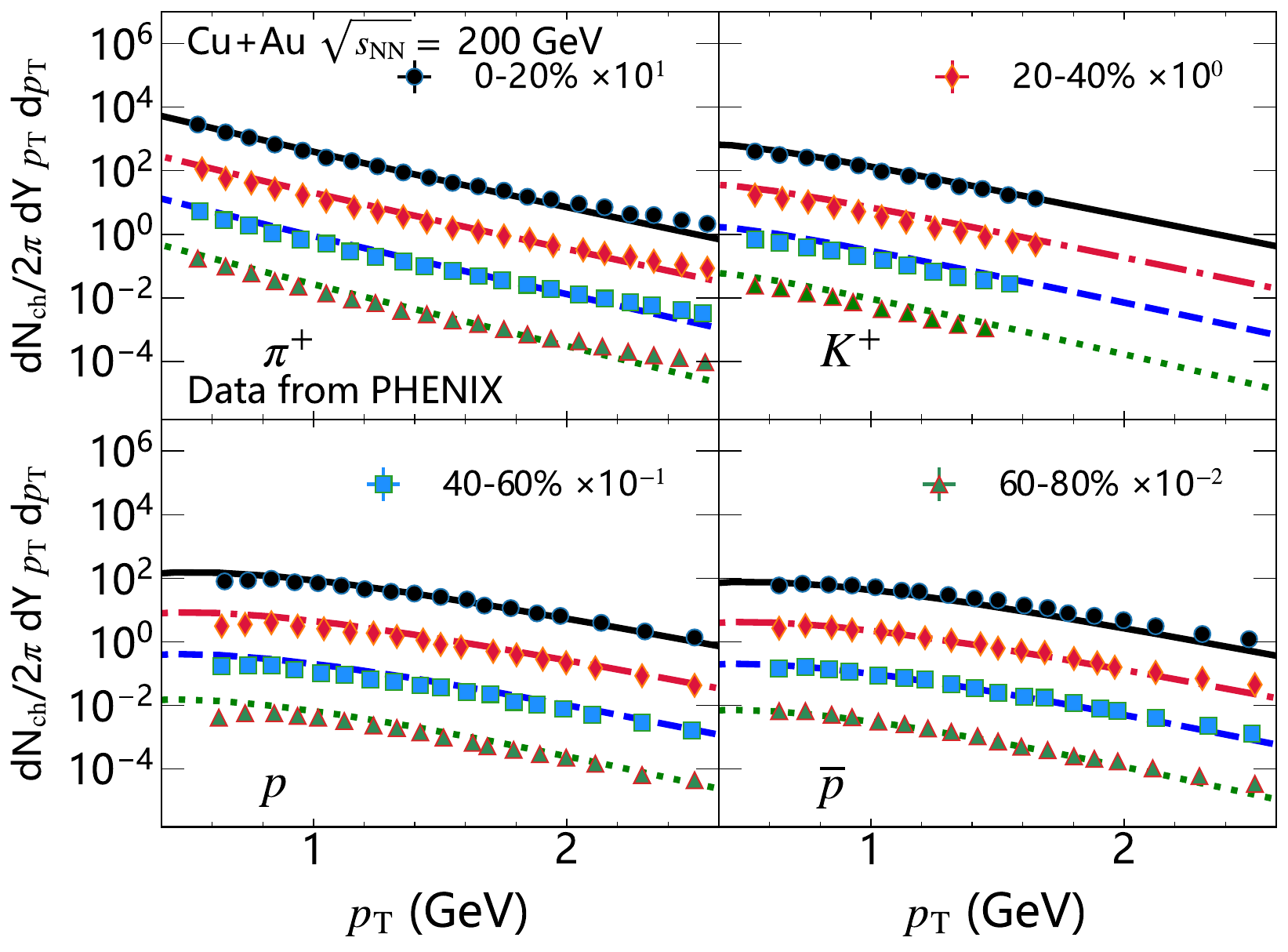} 
\end{center}
\caption{(Color online) The transverse momentum spectra of identified charged hadrons ($\pi^+$, $K^+$, $p$ and $\bar{p}$) at mid-rapidity ($|y|<0.25$) in different centrality bins of Cu+Au collisions at $\sqrt{s_\text{NN}}= 200$~GeV, compared to the PHENIX data~\citep{PHENIX:2023kax}.}
\label{f:particle_spectra}
\end{figure}

We start with validating our model setup by comparing the transverse momentum spectra of identified hadrons between our calculation and the PHENIX data~\citep{PHENIX:2023kax}. 
As shown in Fig.~\ref{f:particle_spectra}, the combination of our initial condition model and the CLVisc hydrodynamic evolution provides a reasonable description of the $p_\text{T}$ spectra of various identified particles ($\pi^+$, $K^+$, $p$ and $\bar{p}$) in Cu+Au collisions at $\snn=$200~GeV. The comparison in the most central collisions helps determine our model parameters listed in the previous section. And using these parameters, the model predictions for other centrality bins agree with the data.

It is important to note that, considering more nucleons from Au nuclei participate in collisions than nucleons from Cu nuclei, the center-of-mass of the QGP produced in Cu+Au collisions moves towards the Au-moving direction with a rapidity of~\cite{STAR:2017ykf}
\begin{align}
y_{\text{CM}} \approx\frac{1}{2}\ln(N^{\text{Au}}_{\text{part}}/N^{\text{Cu}}_{\text{part}}), 
\label{eq:ycm}
\end{align}
where $N^{\text{Au}}_{\text{part}}$ and $N^{\text{Cu}}_{\text{part}}$ are the average numbers of participant nucleons from Au and Cu nuclei, respectively. Therefore, to compare the particle spectra produced from a QGP medium in our computational frame with rapidity $y\approx 0$ to experimental data, a shift of $y_{\text{CM}}$ has been included.

\begin{table}[h]
\centering
\vline
\caption{Values of the center-of-mass rapidity of QGP in Cu+Au collisions at $\sqrt{s_\text{NN}}= 200$~GeV for various centrality regions.}
\begin{tabular}{c c c c c c c c c}
\hline\hline
Centrality & 0-10\%   & 0-20\% & 10-40\% & 20-40\% & 40-60\% & 60-80\% \\ \hline
$y_{\text{CM}}$  & 0.34   & 0.30  & 0.20 & 0.18 & 0.14   & 0.11   \\ \hline\hline
\end{tabular}
\label{table:ycm}
\end{table}

In Tab.~\ref{table:ycm}, we list the $y_{\text{CM}}$ values calculated using the TRENTo model for different centrality regions of Cu+Au collisions at $\sqrt{s_\text{NN}}= 200$~GeV. One can observe an increase of this center-of-mass rapidity as centrality decreases due to a stronger imbalance between participant nucleons from Au and Cu nuclei in more central collisions.

\begin{figure}[tbp!]
\begin{center}
\includegraphics[width=0.85\linewidth]{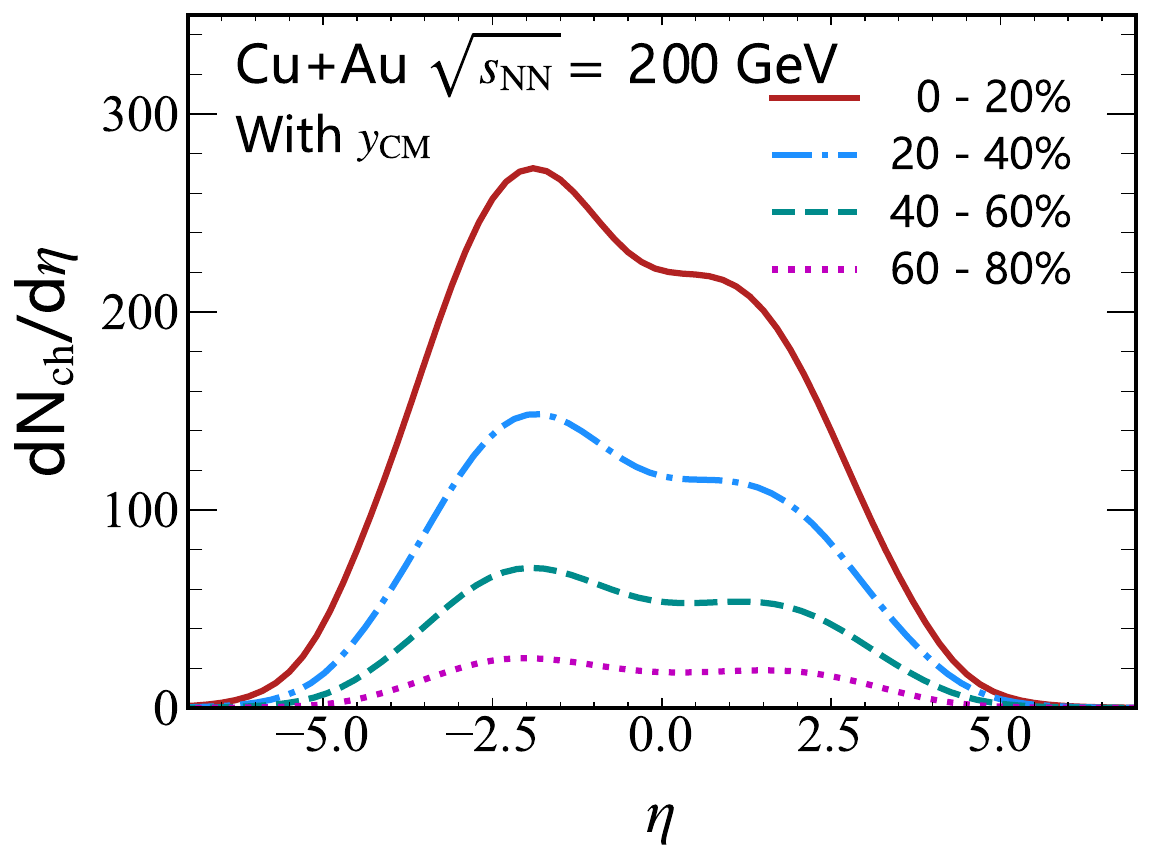} 
\end{center}
\caption{(Color online) Charged particle multiplicity per unit pseudorapidity as a function of pseudorapidity in Cu+Au collisions at $\sqrt{s_\text{NN}}= 200$~GeV, compared between different centrality bins.}
\label{f:dndeta}
\end{figure}

For a better illustration of this center-of-mass shift, we present in Fig.~\ref{f:dndeta} our predictions for the charged hadron yields per unit pseudorapidity ($d\text{N}_{\text{ch}}/d\eta$) as a function of pseudorapidity in Cu+Au collisions at $\snn=$200 GeV. From the figure, we see an increasing particle yield towards more central collisions. Meanwhile, more particles are produced towards the Au-moving ($-\eta$) direction than the Cu-moving ($+\eta$) direction. The center-of-mass rapidity of the system, represented by the dip (or inflection point) in the middle of each curve, locates at the Au-moving side and becomes larger in magnitude in more central collisions. The results shown in this subsection are calculated using our default parameters $H_\text{t}=0.8b$/fm for the initial tilted geometry and $f_v=0.05$ for the initial longitudinal flow velocity. These two parameters do not have a visible effect on the particle yields and spectra here, though they will significantly affect the directed flow coefficient and hyperon polarization later.

\subsection{Collective flows}
\label{section3_sub2}

In this subsection, we first study the directed flow coefficient of charged hadrons produced in Cu+Au collisions at $\sqrt{s_\text{NN}}= 200$~GeV.
The directed flow coefficient at a given rapidity is calculated as the first order Fourier component of the azimuthal angle distribution:
\begin{equation}
\begin{aligned}
v_{1}(y)=\langle\cos(\phi)\rangle=\frac{\int\cos(\phi)\frac{dN}{dy d\phi}d\phi}{\int\frac{dN}{dy d\phi}d\phi},
\label{eq:v1}
\end{aligned}
\end{equation}
where $\phi$ is the azimuthal angle with respect to the second order participant plane in this work.
As discussed in Sec.~\ref{section2_sub1}, we use a smooth initial condition averaged over 5000 events whose second order participant planes are aligned with the $+\hat{x}$ direction. Therefore, the participant plane of the smooth initial condition coincides with the second order event plane of the final state hadrons within our setup. According to Ref.~\cite{STAR:2017ykf}, the directed flow coefficient evaluated based on the second order event plane is comparable to the directed flow coefficient obtained using the 3-particle correlation method ($v_1\{3\}$) in experiments. On the other hand, since we have not extracted the information of spectators in each event and event-by-event fluctuation is not taken into account in this work, our result cannot be quantitatively compared to the directed flow coefficient obtained using the spectator plane method ($v_1\{\text{SP}\}$) in experiments yet.

\begin{figure}[tbp!]
\begin{center}
\includegraphics[width=0.85\linewidth]{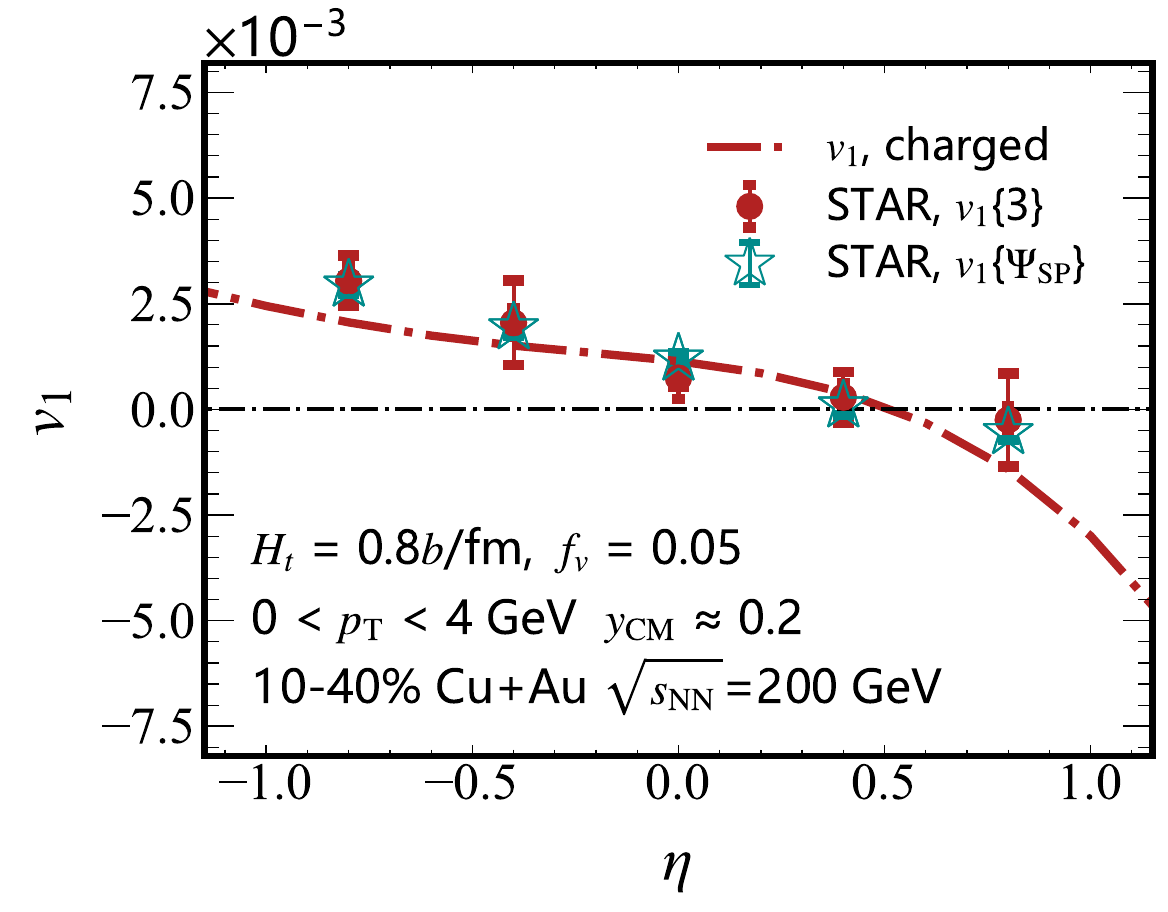} \\
\includegraphics[width=0.85\linewidth]{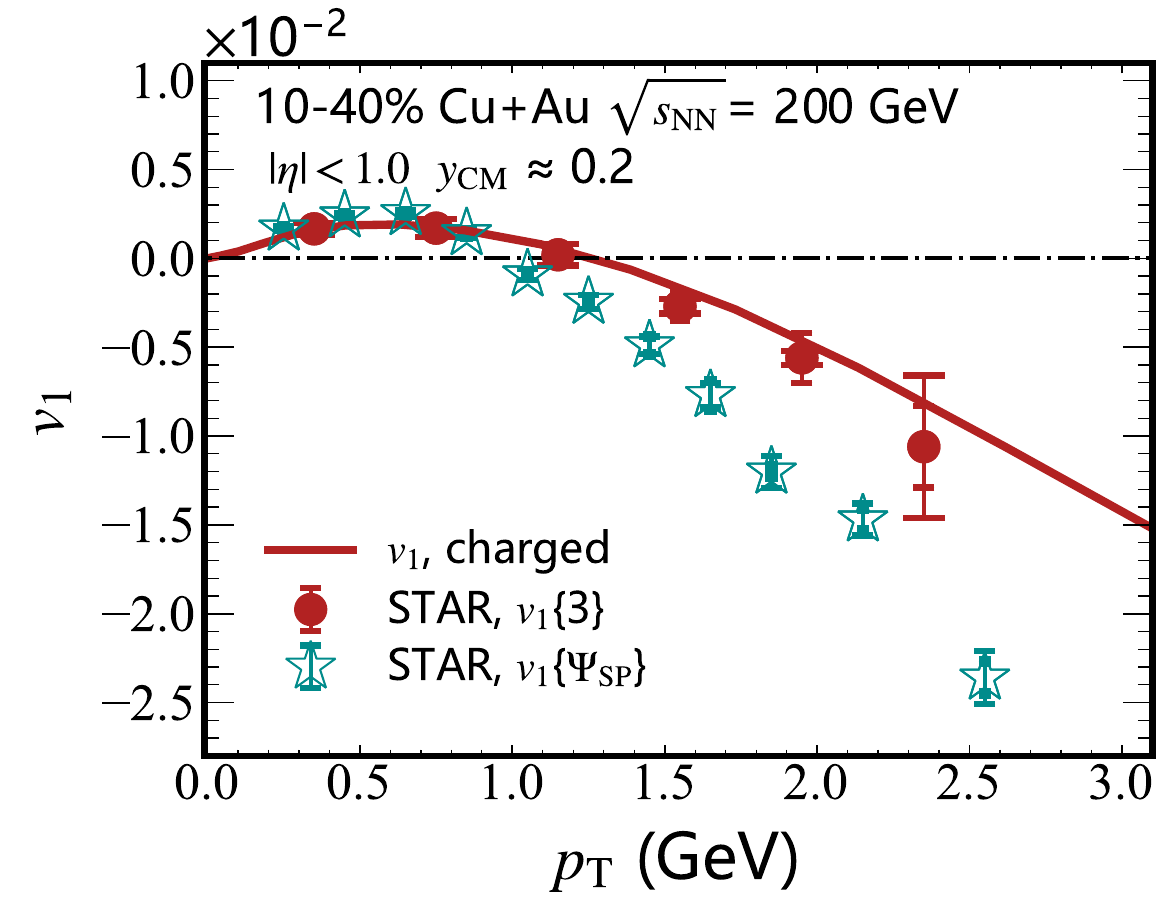} \\
\includegraphics[width=0.85\linewidth]{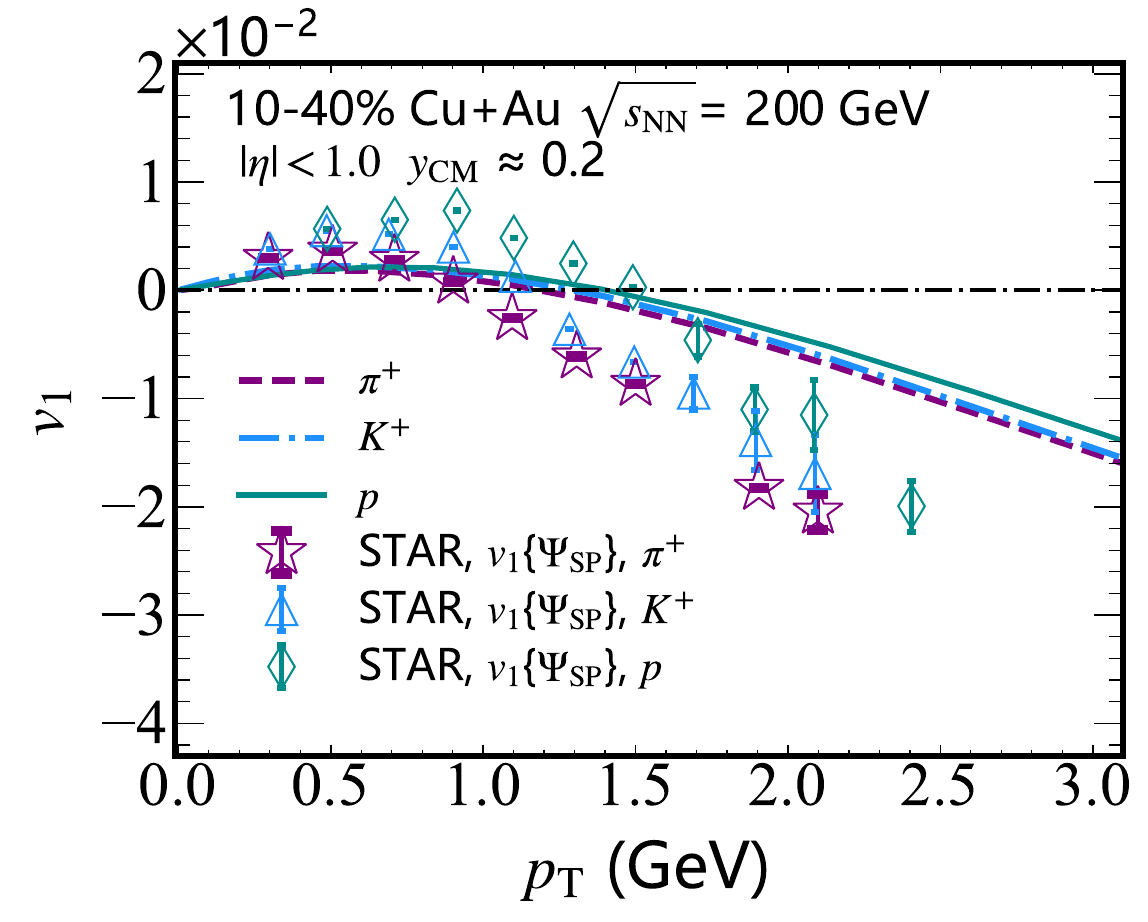} 
\end{center}
\caption{(Color online) The directed flow of hadrons in 10-40\% Cu+Au collisions at $\snn=200$~GeV, upper panel for the pseudorapidity dependence of the charged particle $v_1$, middle and lower panels for the transverse momentum dependence of $v_1$ of charged particles and identified particles, respectively. Results from our model calculation are compared to the STAR data~\citep{STAR:2017ykf}.}
\label{f:v1}
\end{figure}

In the upper panel of Fig.~\ref{f:v1}, we first present $v_1$ of charged particles as a function of pseudorapidity in 10-40\% Cu+Au collisions at $\snn=200$ GeV. 
By using the linear ansatz of $H_\mathrm{t}= 0.80 b$/fm between the tilted geometry parameter and the impact parameter, and a constant longitudinal flow parameter $f_{v}=0.05$ in the initial state, we obtain a satisfactory description of the STAR data~\citep{STAR:2017ykf}. The pseudorapidity dependence of $v_{1}$ shows a negative slope around the middle $\eta$ region. Furthermore, it exhibits an asymmetric structure between forward and backward rapidities, with larger value in the backward region than in the forward region. Because of the imbalanced energy deposition from Au and Cu nuclei, $v_1$ crosses zero in the forward rapidity region. Note that the pseudorapidity dependence of $v_{1}$ here is similar to a mirror image of $\varepsilon_1$ shown in Fig.~\ref{f:cuau200ecc1} with respect to the $\eta=0$ axis, when $H_\text{t}$ has the same setup. This implies the tilted initial geometry of the QGP qualitatively determines the pattern of $v_1$ through hydrodynamic expansion. Later in Fig.~\ref{f:v1_depen} we will analyze effects of different components of our initial condition model on $v_1$ in detail. For transverse momentum ($p_\mathrm{T}$) integrated $v_1$ as a function of $\eta$, there is no apparent difference between experimental data measured using the 3-particle correlation method and the spectator plane method.

In the middle panel of Fig.~\ref{f:v1}, we show the $p_\mathrm{T}$ dependence of $v_1$ of charged particles within the $|\eta|<1$ region for 10-40\% Cu+Au collisions at $\snn=200$~GeV. At large $p_\text{T}$, the experimental data measured using the two methods deviate from each other, and our result is consistent with the one obtained using the 3-particle correlation method ($v_{1}\{3\}$). The directed flow undergoes a sign change near $p_{\text{T}}=1$~GeV, which can be understood with the momentum conservation in the transverse plane~\cite{Teaney:2010vd}. Since more low $p_\text{T}$ particles are emitted from the QGP than higher $p_\text{T}$ ones, the former determine the sign of the $p_\text{T}$-integrated $v_1$. As shown in the upper panel, the $p_\mathrm{T}$-integrated $v_1$ is positive around mid-rapidity of Cu+Au collisions, and therefore, we have $v_1>0$ for low $p_\mathrm{T}$ particles here, but $v_1<0$ for high $p_\mathrm{T}$ ones. This sign reversal of $v_1$ in Cu+Au collisions can also be seen in the PHENIX data~\cite{PHENIX:2015zbc}.

In the lower panel of Fig.~\ref{f:v1}, we study the particle species dependence of $v_1$. At low $p_\text{T}$ ($<0.5$~GeV), no apparent difference between $\pi^{+}$, $K^{+}$, and $p$ is observed. However, at higher $p_\text{T}$, one can clearly see that the directed flow coefficients of $\pi^{+}$ and $K^{+}$ decrease faster than that of $p$, partly due to the mass effect on transferring anisotropic flow from the hydrodynamic medium to different species of hadrons. Although our result qualitatively agrees with the data, the splitting between $\pi^{+}$, $K^{+}$, and $p$ from our calculation appears smaller than that shown in the data. This could result from the ignorance of hadronic scatterings and event-by-event fluctuations in our current work, and also the deviation between our analysis of $v_1$ based on the event plane method and the experimental data based on the spectator plane method.

\begin{figure}[tbp!]
\includegraphics[width=0.85\linewidth]{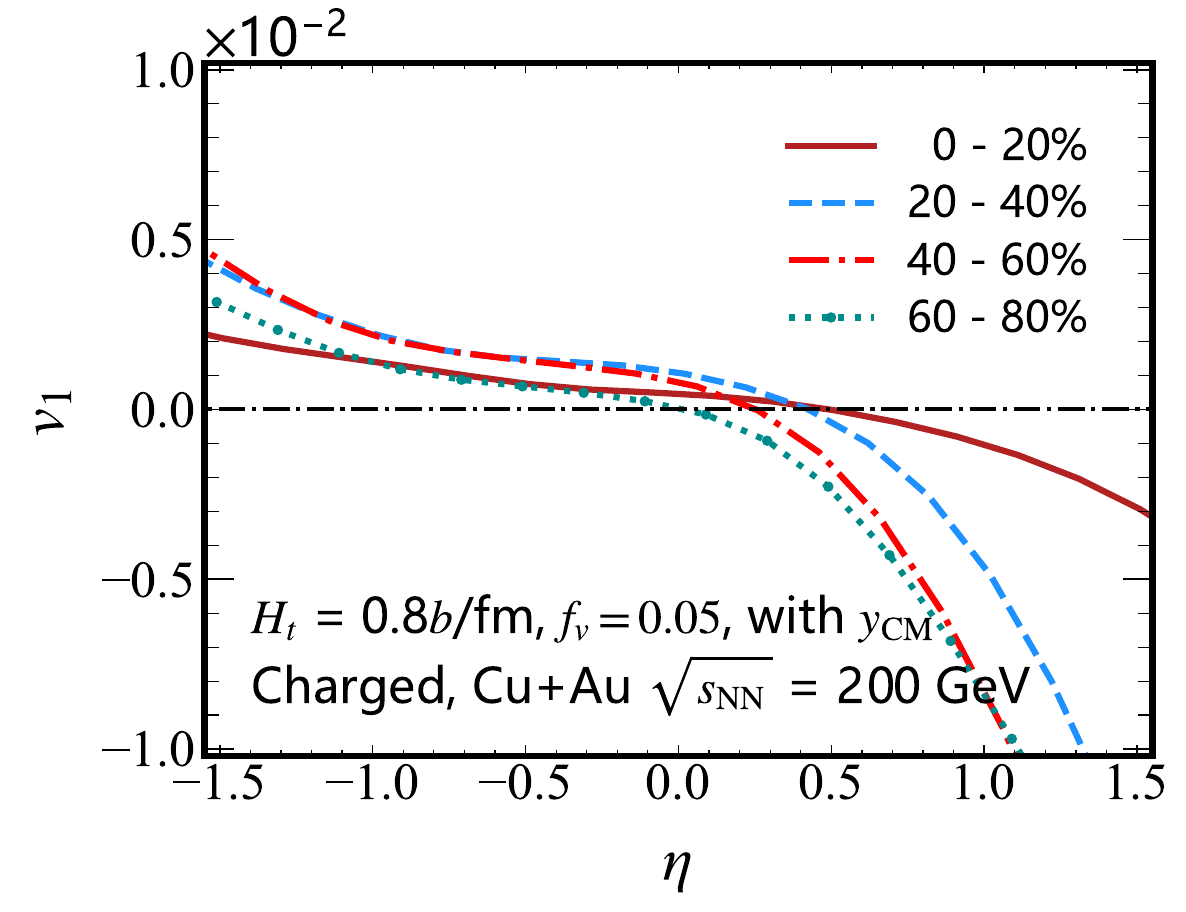}
\includegraphics[width=0.85\linewidth]{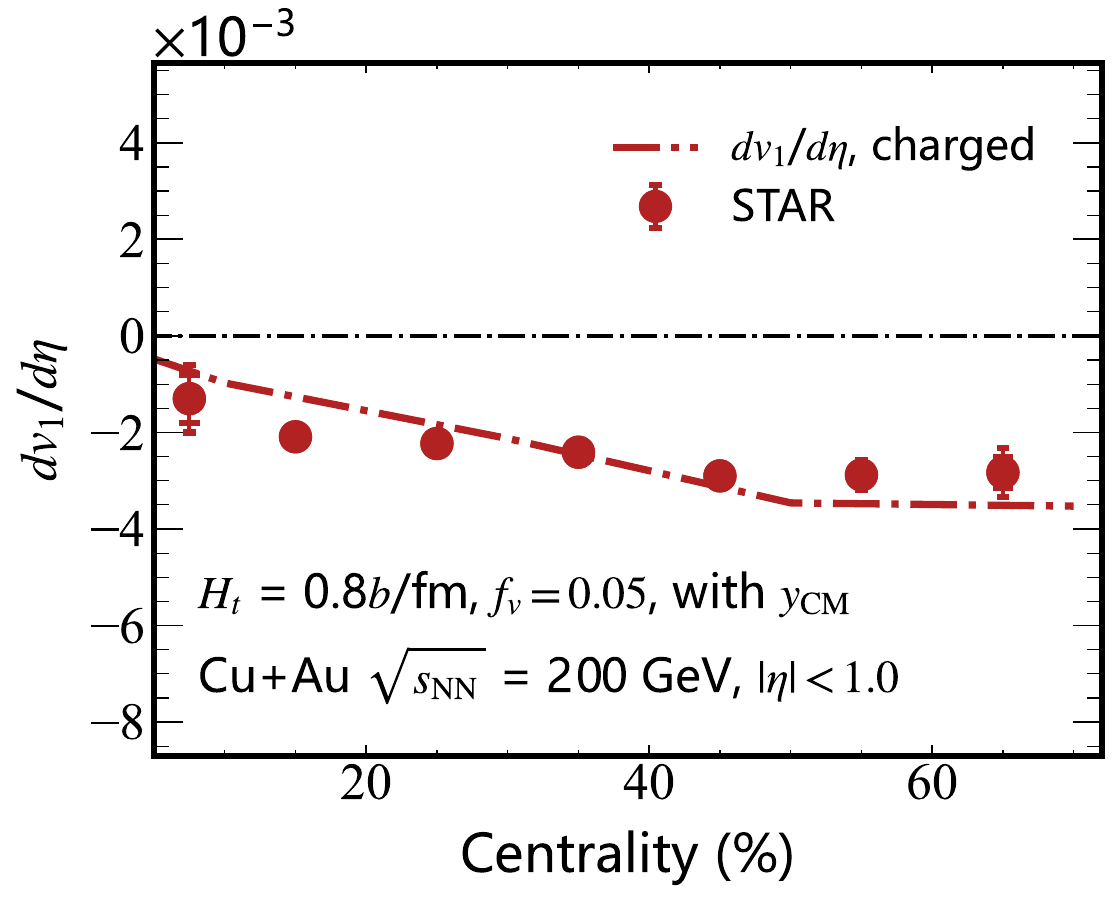}
\caption{(Color online) Centrality dependence of $v_1$ in Cu+Au collisions at $\snn=200$~GeV, upper panel for $v_1(\eta)$ compared between different centrality bins, lower panel for the slope of $v_1(\eta)$ within $|\eta|<1$ as a function of centrality. The experimental data are taken from the STAR Collaboration~\cite{STAR:2017ykf}.}
\label{f:v1_cen}
\end{figure}

In Fig.~\ref{f:v1_cen}, we investigate the centrality dependence of the charged hadron $v_1$. In the upper panel, we present $v_1$ as a function of $\eta$ for various centrality bins. The magnitude of $v_1$ and the position of $\eta$ where $v_1$ crosses zero depends on the interplay between the tilted geometry of the medium, the size of the medium, and the asymmetric energy deposition from Cu and Au nuclei. As the impact parameter (or centrality) increases, the medium becomes more tilted, causing stronger asymmetry of the medium between $+\hat{x}$ and $-\hat{x}$ directions in the forward and backward rapidity regions, and therefore larger magnitude of $v_1$. However, the size of the QGP decreases, which can lead to smaller magnitude of $v_1$, especially in very peripheral collisions. Meanwhile, in more peripheral collisions, the imbalanced energy deposition from Cu and Au nuclei decreases. Therefore, the crossing point for $v_1=0$, as shown in the upper panel of Fig.~\ref{f:v1}, should move towards zero as centrality increases. On the other hand, the center-of-mass shift towards the negative $\eta$ direction, shown in Tab.~\ref{table:ycm} is also reduced. Overall, we see the crossing point approaches zero from central to peripheral collisions.
    
To quantify the decreasing trend of $v_1$ with respect to $\eta$, we present the slope of the $v_1(\eta)$ function in the lower panel of Fig.~\ref{f:v1_cen}, as a function of centrality. For each centrality bin, the slope parameter is extracted from a linear function fit to the values of $v_1$ within $|\eta|<1$. We see the magnitude of the this slope parameter increases with centrality, indicating a stronger tilt of the QGP fireball in more peripheral collisions. Result from our calculation agrees well with the STAR data~\cite{STAR:2017ykf}. 

\begin{figure}[tbp!]
\includegraphics[width=0.85\linewidth]{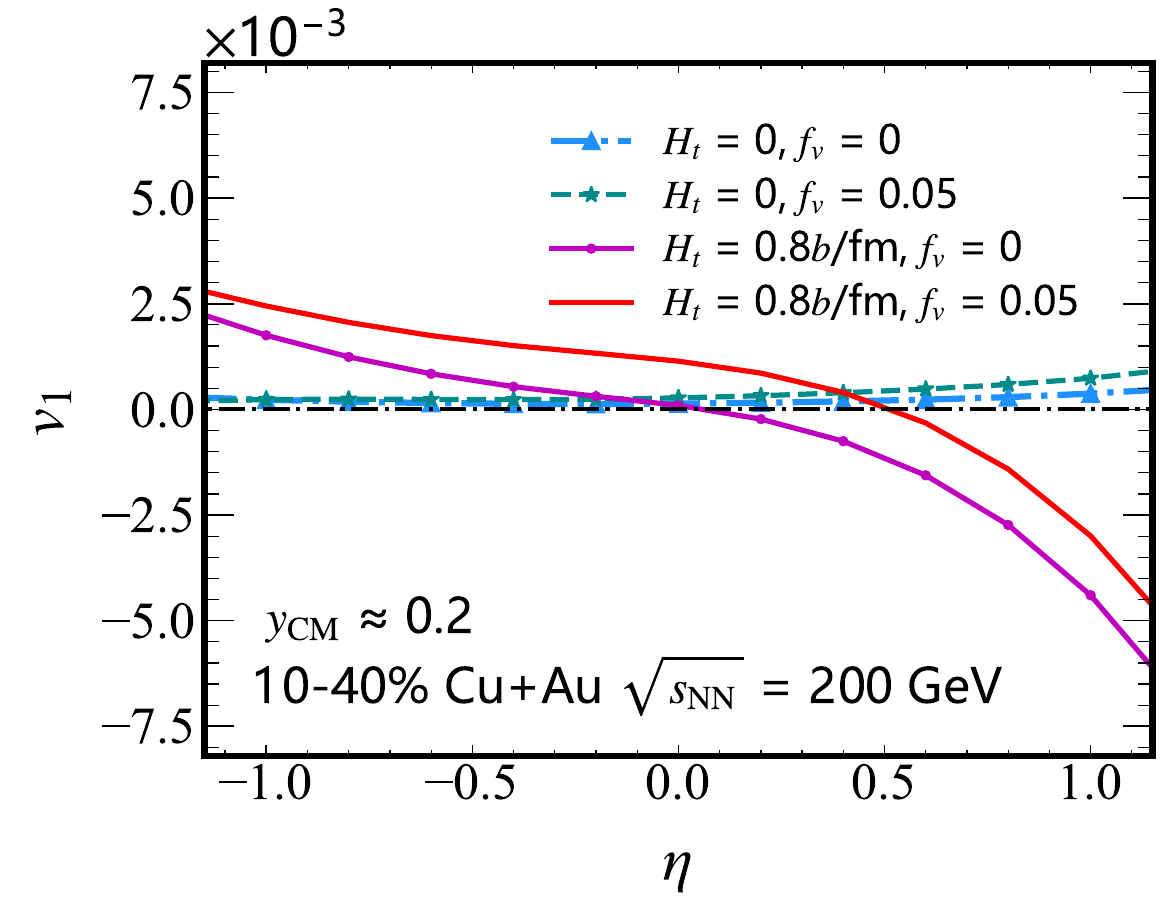}
\includegraphics[width=0.85\linewidth]{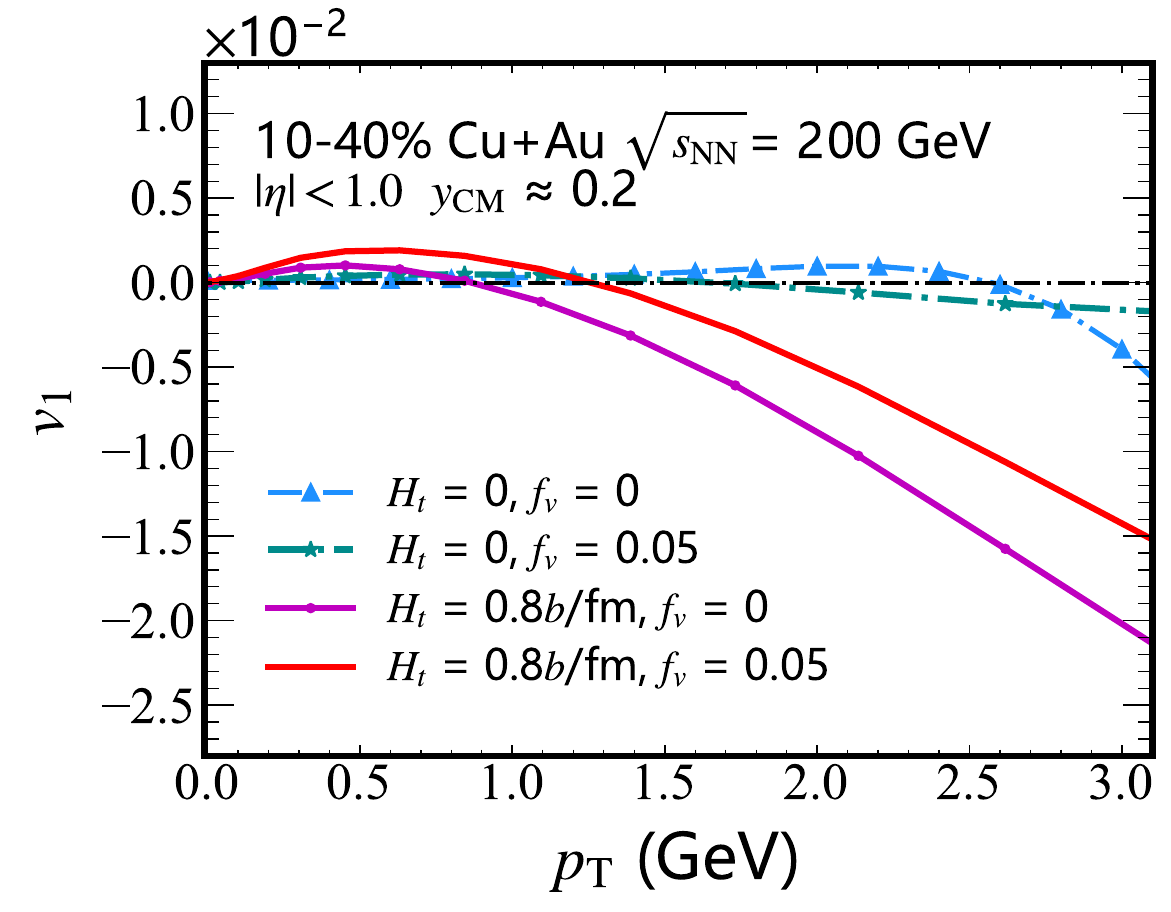}
\caption{(Color online) Comparison of the charged particle $v_1$ between different parameters of our initial condition model for 10-40\% Cu+Au collisions at $\snn=200$~GeV, upper panel for $p_\mathrm{T}$-integrated $v_1$ as a function of $\eta$, lower panel for $v_1$ as a function of $p_\mathrm{T}$ within $|\eta|<1$.}
\label{f:v1_depen}
\end{figure}

In Fig.~\ref{f:v1_depen}, we further investigate effects of various model components on the directed flow coefficient. In the upper panel, we present the $\eta$ dependence of $v_1$, compared between different values of $H_\mathrm{t}$ and $f_v$. We find that when $H_\text{t}=0$ and $f_{v}=0$, the asymmetric energy deposition from Cu and Au nuclei only causes a mild positive value of $v_1$. As discussed in Fig.~\ref{f:cuau200ecc1}, this is due to the larger pressure gradient on the Cu side than on the Au side. The slope parameter ($dv_1/d\eta$) is close to zero. Note that in symmetric (e.g. Au+Au) collisions, $v_1(\eta)$ should be zero with this setup. When taking into account the tilted geometry of the QGP, i.e., $H_\text{t}\ne0$, much stronger asymmetry is produced in the transverse plane between $+\hat{x}$ and $+\hat{x}$ directions, as previously shown in Fig.~\ref{f:cuau200ecc1}. This leads a strong directed flow of final state hadrons, which decreases with pseudorapidity. It worths noticing that unlike the positive correlation between higher order eccentricities and harmonic flow coefficients, the correlation between $\varepsilon_1$ and $v_1$ depends on its source. For asymmetry directly introduced to the transverse plane, as the one caused by the imbalanced energy deposition from Cu and Au nuclei between $+\hat{x}$ and $-\hat{x}$ directions, $\varepsilon_1$ and $v_1$ are positively correlated with each other through the hydrodynamic expansion in the transverse plane, similar to higher order asymmetries. In contrast, the transverse asymmetry introduced via the tilted QGP geometry with respect to the longitudinal direction is more complicated. The hydrodynamic expansion along the longitudinal direction continuously change the asymmetry in transverse planes at different rapidities, and the medium becomes less tilted as it evolves. Therefore, the final state $v_1$ is not necessarily positively correlated with the initial state $\varepsilon_1$. As demonstrated in our earlier work~\cite{Jiang:2021ajc}, the average pressure gradient in the transverse plane generated by this tilted geometry actually points towards the direction opposite to $\vec{\mathcal{E}}_{1}$, leading to a negative correlation between $\varepsilon_1$ and $v_1$ in the end. By comparing Fig.~\ref{f:cuau200ecc1} and the upper panel of Fig.~\ref{f:v1_depen}, one can see different correlations between $\varepsilon_1$ and $v_1$ when different initial conditions are applied. In symmetric Au+Au collisions, the tilted deformation of QGP should generate an odd function of $v_1$ with respect to $\eta$~\cite{Jiang:2021ajc}. Here, since the energy deposition from Cu and Au nuclei is asymmetric along both the impact parameter ($x$) direction and the longitudinal ($z$) direction, additional imbalance can be observed in the magnitude of $v_1$ between the forward and backward directions. Our result with $H_\text{t}\ne0$ but $f_v=0$ is similar to that obtained using the AMPT model~\cite{Chen:2005zy}. Moreover, incorporating the initial longitudinal flow ($f_v\ne 0$) further modifies the transverse asymmetry at different rapidities. As illustrated in Refs.~\cite{Csernai:2011gg, Shen:2020jwv}, such longitudinal flow field drives the upward moving matter towards the forward direction, while downward moving matter backward. Here, we see this longitudinal flow causes an upward shift of the $v_1(\eta)$ function, resulting in a shift of the crossing point of $v_1(\eta)=0$ towards larger $\eta$.

\begin{figure}[tbp!]
\includegraphics[width=0.85\linewidth]{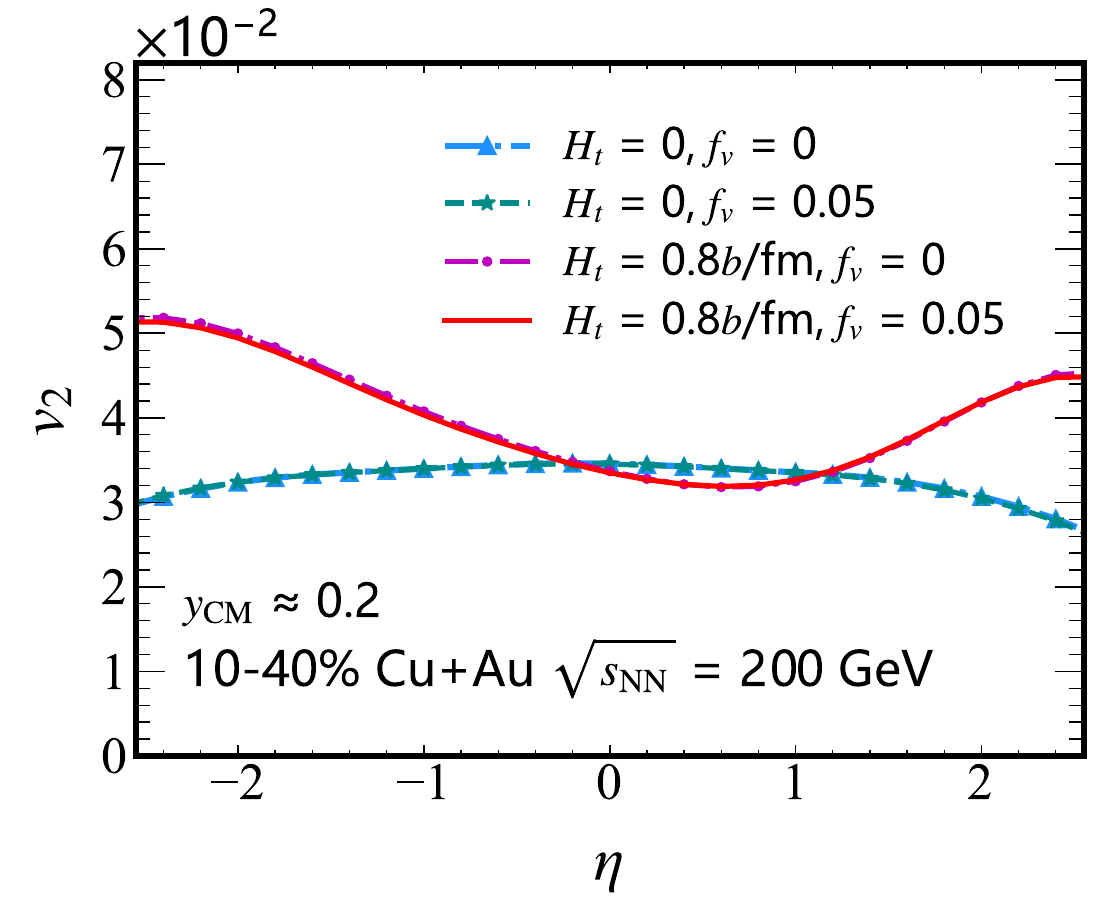}
\caption{(Color online) Elliptic flow coefficient of charged particles as a function of pseudorapidity in 10-40\% Cu+Au collisions at $\snn=200$~GeV, compared between different parameter setups of our initial condition model.}
\label{f:v2_eta}
\end{figure}

Similar conclusions can be drawn from the $p_\text{T}$ dependence of $v_1$ in the lower panel of Fig.~\ref{f:v1_depen}.
When $H_\text{t}=0$ and $f_v=0$, the directed flow originating from the asymmetric energy deposition between Cu and Au nuclei is small within the $|\eta|<1$ region. Introducing the tilted geometry ($H_\mathrm{t}\ne 0$) significantly enhances the magnitude of $v_1$. According to our previous discussion for the middle panel of Fig.~\ref{f:v1}, positive value of $p_\mathrm{T}$-integrated $v_1$ leads to positive $v_1$ at low $p_\mathrm{T}$ while negative $v_1$ at high $p_\mathrm{T}$. Adding the initial longitudinal flow ($f_v\ne 0$) further enhances the positive $v_1$ around mid-rapidity, which increases the $v_1$ of low $p_\mathrm{T}$ particles and meanwhile shifts the crossing point of $v_1(p_\mathrm{T})=0$ to higher $p_\mathrm{T}$.

\begin{figure}[tbp!]
\includegraphics[width=0.9\linewidth]{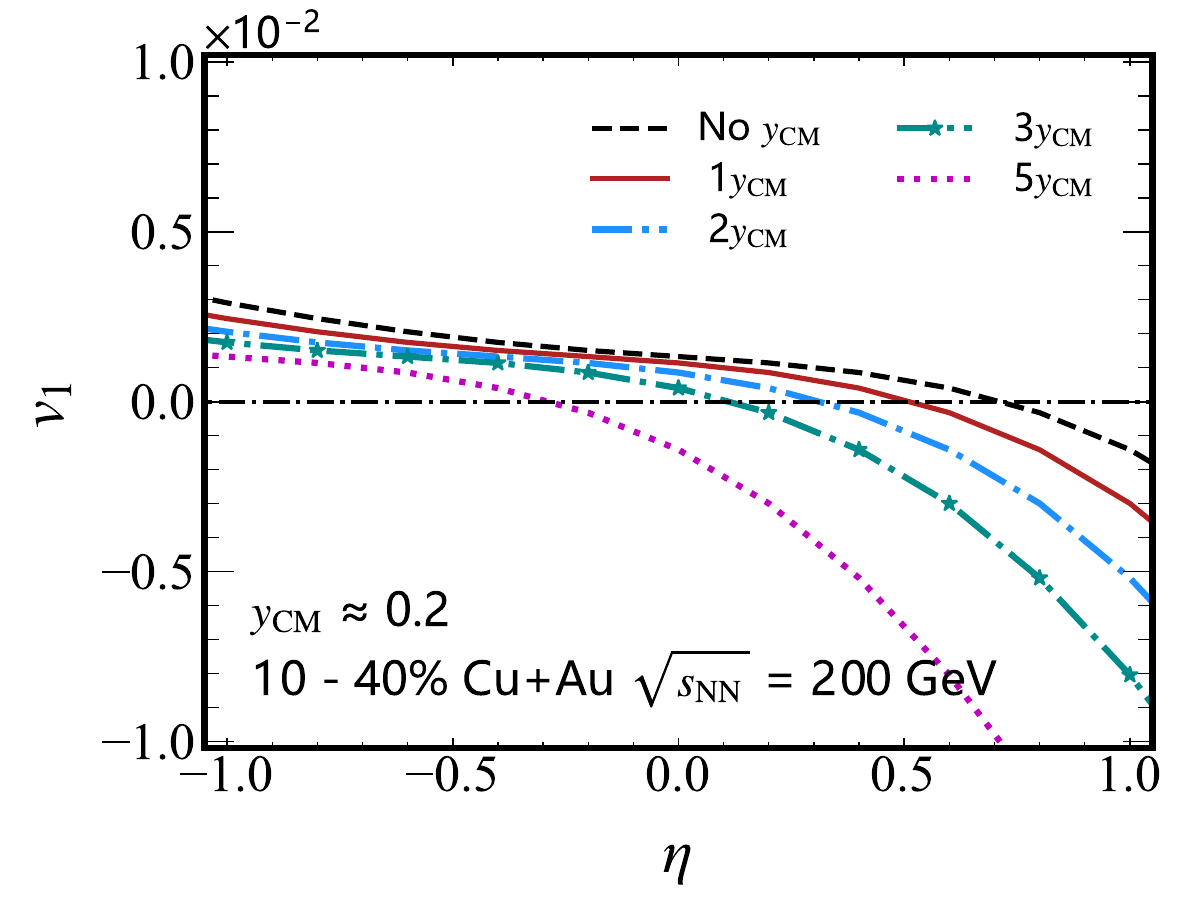} \\
\includegraphics[width=0.9\linewidth]{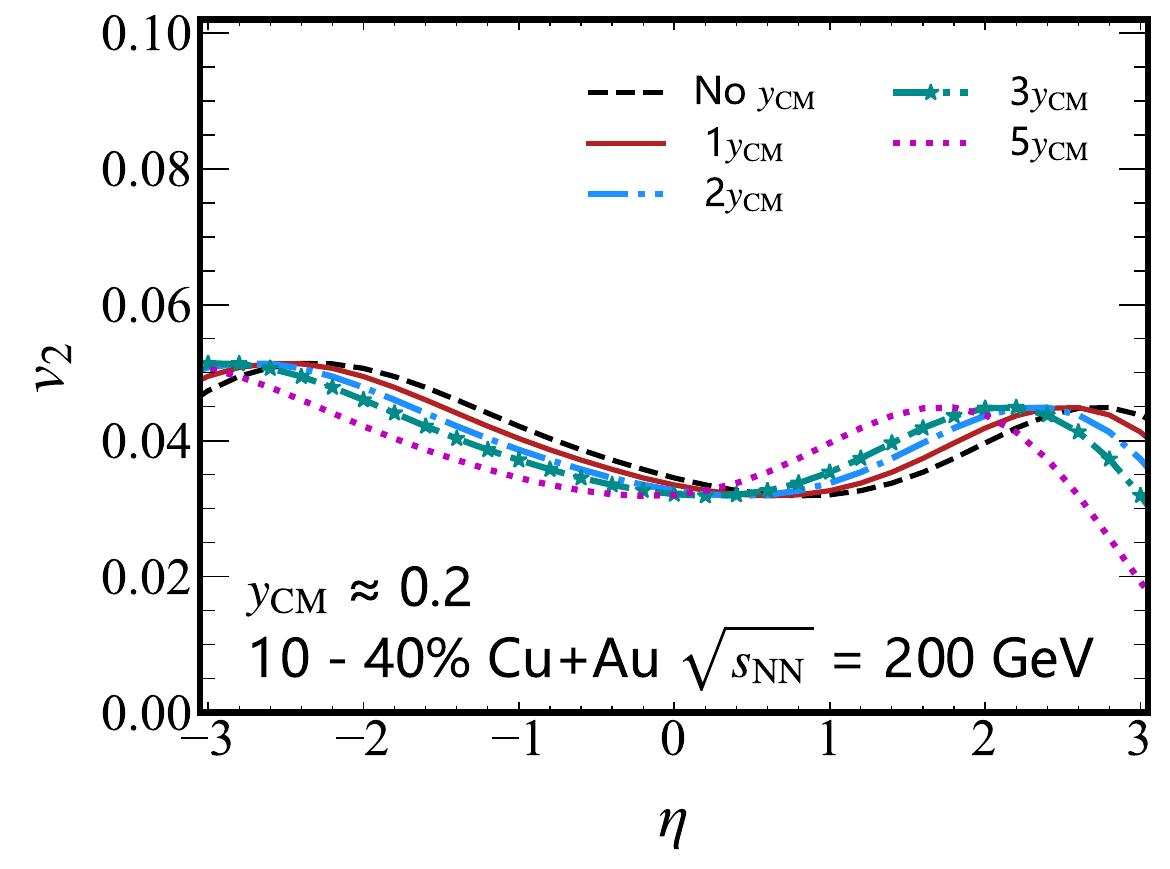}
\caption{(Color online) Effect of the center-of-mass shift on the rapidity dependence of collective flow coefficients in 10-40\% Cu+Au collisions at $\snn=200$~GeV, upper panel for directed flow and lower panel for elliptic flow.}
\label{f:v1_ycm}
\end{figure}

\begin{figure*}[tbp!]
\includegraphics[width=0.32\linewidth]{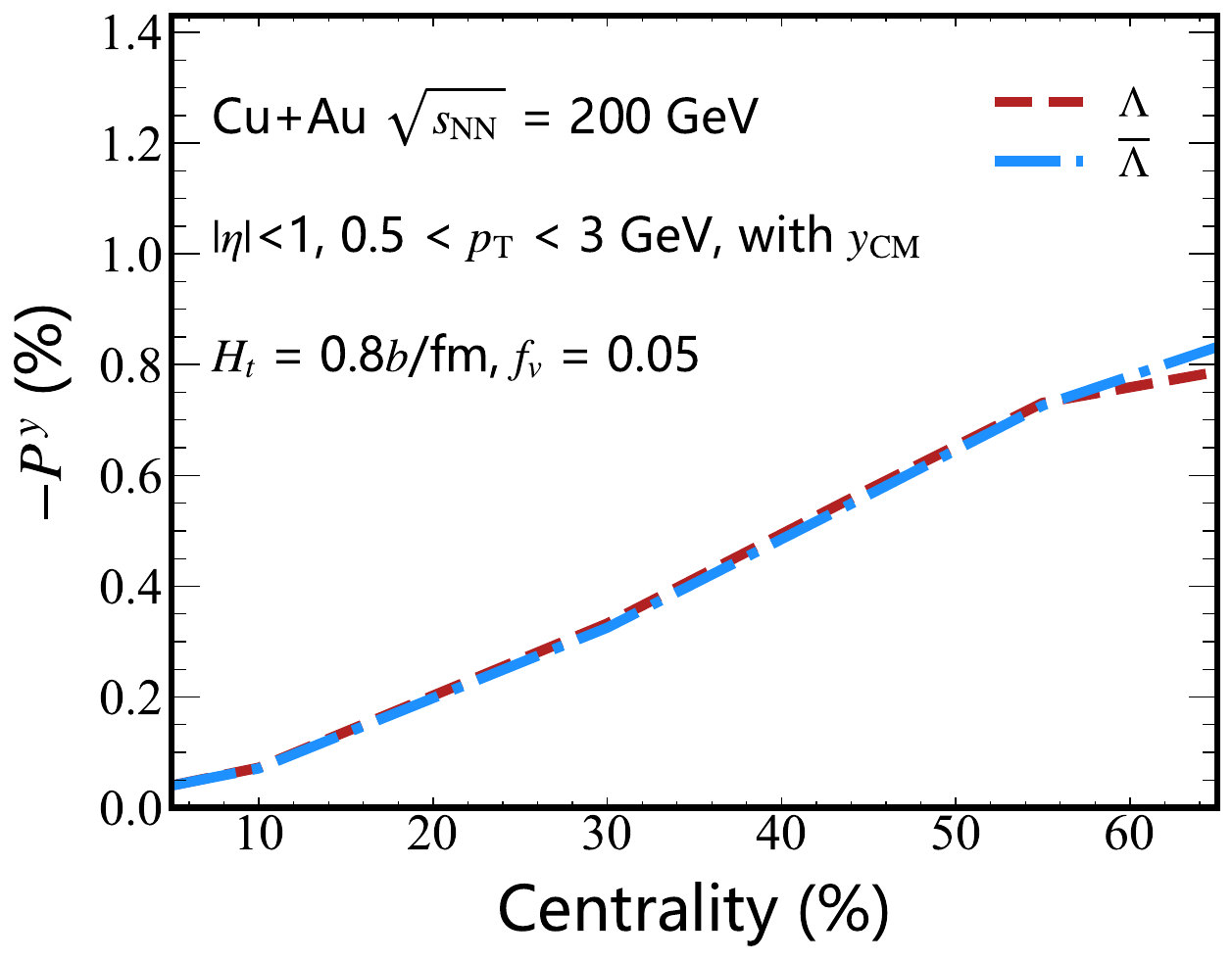}
\includegraphics[width=0.32\linewidth]{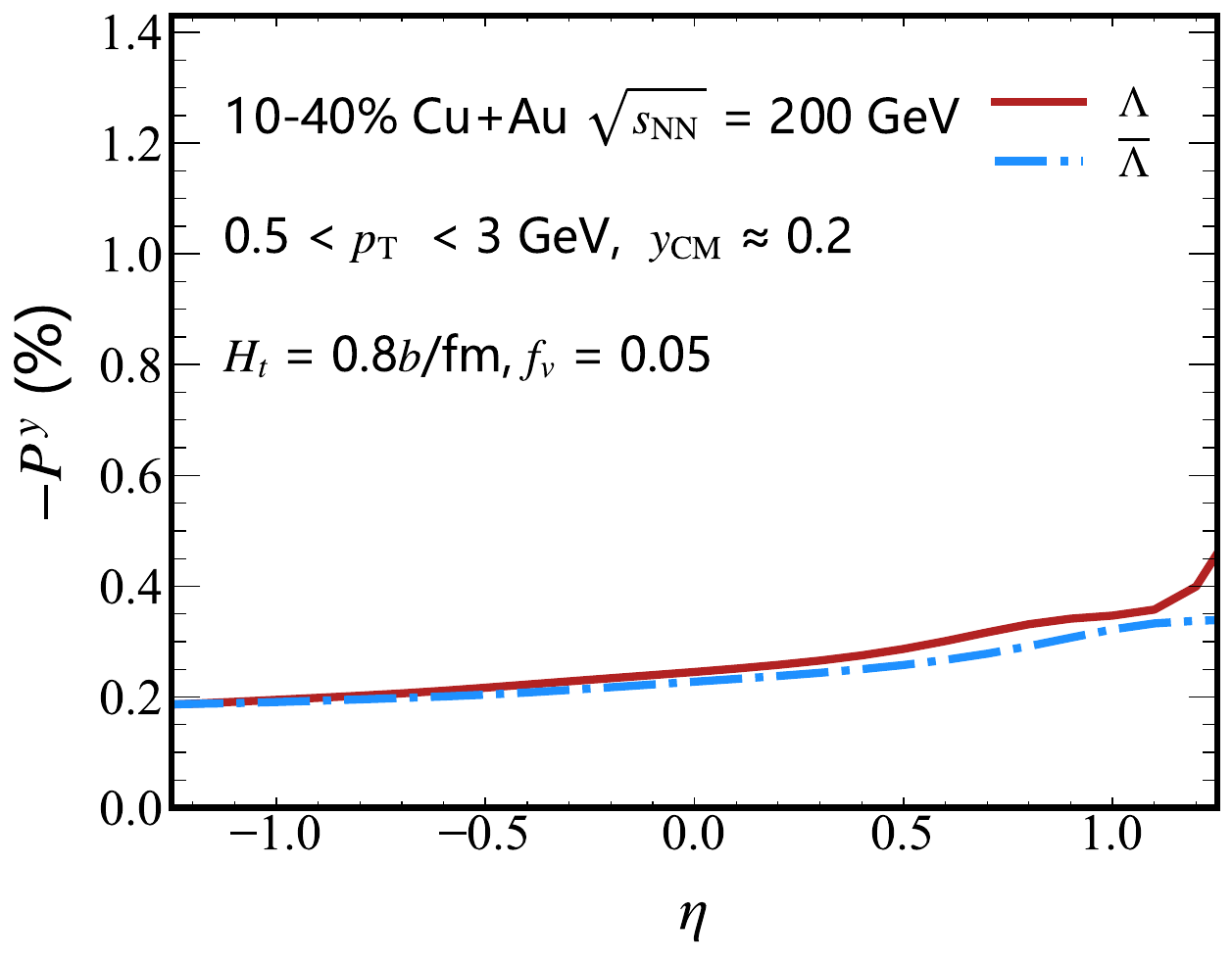}
\includegraphics[width=0.32\linewidth]{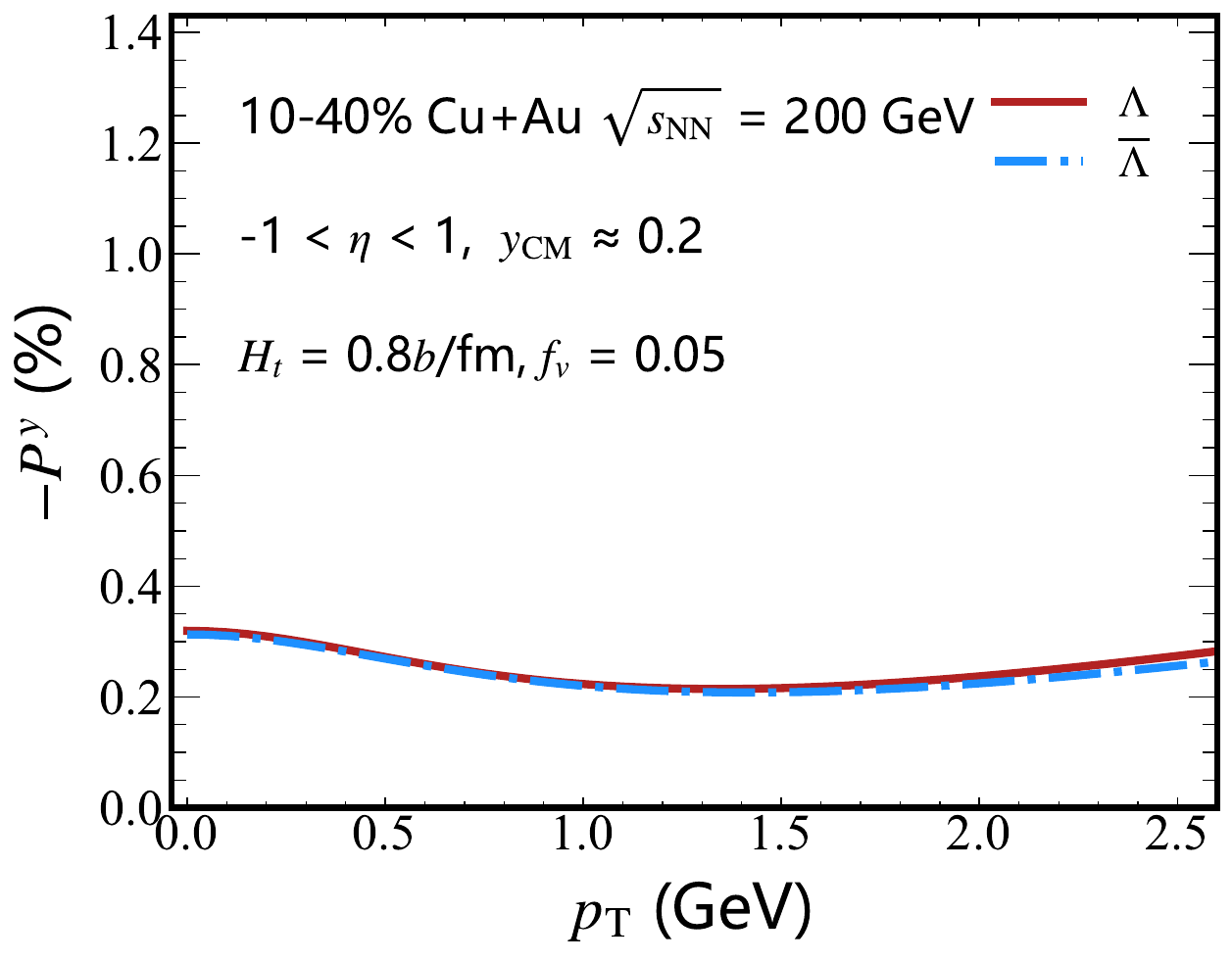}
\caption{(Color online) Global polarization of $\Lambda$ and $\bar{\Lambda}$ hyperons as functions of centrality (left panel), pseudorapidity (middle panel), and transverse momentum (right panel) in 10-40\% Cu+Au collisions at $\snn=200$~GeV.}
\label{f:global_py}
\end{figure*}

In Fig.~\ref{f:v2_eta}, we also study effects of different initial conditions on the elliptic flow coefficient ($v_2$) of charged particles in Cu+Au collisions. It is interesting to note that the asymmetry in the transverse plane caused by the tilted geometry of the QGP has qualitative impact on the pseudorapidity dependence of $v_2$: $H_\text{t}=0.8b$/fm leads to an enhancement of $v_2$ at large forward and backward rapidities, which is absent when $H_\text{t}=0$. Such enhancement at large rapidity was not seen in the AMPT result for Cu+Au collisions~\cite{Chen:2005zy}, probably due to different degrees of the tilted geometry achieved in different models. On the other hand, study based on symmetric (Pb+Pb) collisions~\cite{Csernai:2011gg} pointed out that this enhancement at large rapidity can disappear when the initial state center-of-mass rapidity fluctuation is taken into account. This is also why such enhancement has not been observed in symmetric heavy-ion collision experiments~\cite{PHOBOS:2004vcu,CMS:2012zex,ALICE:2016tlx}. Nevertheless, within our model, we have found that the $v_2$ enhancement at large rapidity in Cu+Au collisions appears much stronger than that in Au+Au collisions. Therefore, it is still of great interest if experiments can measure the rapidity dependence of $v_2$ in Cu+Au collisions and help understand the effect of tilted QGP geometry. Furthermore, in Fig.~\ref{f:v2_eta}, we observe negligible impact of the initial longitudinal flow (value of $f_v$) on the rapidity dependence of $v_2$, which is different from its impact on $v_1$ shown in Fig.~\ref{f:v1_depen}. Thus, a combined investigation on different orders of collective flows will provide a more stringent constraint on the geometry and longitudinal flow profile of the initial state of the QGP.

At the end of this subsection, we demonstrate effects of the center-of-mass shift ($y_{\text{CM}}$) on the rapidity dependence of collective flow coefficients in Fig.~\ref{f:v1_ycm}. In the upper panel, we present $v_1$ as a function of $\eta$, compared between different amount of rapidity shift, varying from zero to five times of $y_{\text{CM}}$ (0.2 for 10-40\% Cu+Au collisions at $\snn=200$~GeV). Since the medium asymmetry along the $x$ direction increases with rapidity when $H_\text{t}=0.8b$/fm and $f_v=0.05$, $v_1$ varies more quickly with respect to $\eta$ at large $\eta$. Hence, a larger rapidity shift of the function $v_1(y)$ towards the Au-moving direction naturally leads to a more steeply falling $v_1$ with $\eta$ within the $|\eta|< 1.0$ region, and meanwhile drives the point at which $v_1$ transitions from positive to negative towards smaller $\eta$. Similarly, in the lower panel, we find this rapidity shift moves the local extrema of $v_2(\eta)$, including two local maxima at forward and backward rapidities and one local minimum around mid-rapidity, towards the Au-moving direction. Therefore, including this rapidity shift is crucial for a quantitative comparison between hydrodynamic calculation and experimental data for asymmetric heavy-ion collisions.

\subsection{Global polarization }
\label{section3_sub3}

Apart from collective flows, spin polarization is another quantity that is sensitive to the geometry and flow velocity profile of the QGP. Therefore, these two groups of observables are intrinsically related to each other and can be investigated within the same hydrodynamic framework~\cite{Voloshin:2017kqp,Ivanov:2020wak,Ryu:2021lnx,Jiang:2023vxp}. In this subsection, we explore the global polarization of $\Lambda$ and $\bar{\Lambda}$ hyperons in Cu+Au collisions.

Following Refs.~\cite{Yi:2021ryh,Wu:2022mkr,Yi:2023tgg,Becattini:2013fla,Fang:2016vpj}, we employ a modified Cooper-Frye formalism to calculate the polarization pseudovector of spin-1/2 fermions as follows,
\begin{equation}
\mathcal{S}^{\mu}(\vec{p})=\frac{\int d \Sigma \cdot p \mathcal{J}_{5}^{\mu}(p, X)}{2 m \int d \Sigma \cdot \mathcal{N}(p, X)},
\end{equation}
where $\Sigma^\mu$ denotes the freezeout hypersurface, $\mathcal{J}^{\mu}_5$ is the axial charge current density and $\mathcal{N}^{\mu}$ is the fermion number density in the phase space $(p,X)$, and $m$ is the fermion mass.
According to the decomposition of the vector product between the thermal vorticity tensor and the 4-momentum vector~\citep{Karpenko:2018erl} or the quantum kinetic theory~\citep{Yi:2021ryh,Hidaka:2017auj,Yi:2021unq}, $\mathcal{S}$ can be decomposed into five terms: the thermal vorticity ($\mathcal{S}_{\textrm{thermal}}^{\mu}$), the shear tensor ($\mathcal{S}_{\textrm{shear}}^{\mu}$),
the fluid acceleration minus temperature gradient ($\mathcal{S}_{\textrm{accT}}^{\mu}$),
the gradient of chemical potential over temperature ($\mathcal{S}_{\textrm{chemical}}^{\mu}$),
and the external electromagnetic field ($\mathcal{S}_{\textrm{EB}}^{\mu}$).
Here, $S^\mu_{\textrm{shear}}$ and $S^\mu_{\textrm{chemical}}$ are also known as the shear-induced polarization (SIP) and the baryonic spin Hall effect (SHE)~\cite{Liu:2020dxg}.
Since the external electromagnetic field decays very fast and its evolution has not been well constrained in heavy-ion collisions yet, we only include the first four terms in this work but neglect $S^{\mu}_\mathrm{EB}$.
Detailed expressions of these terms can be found in Refs.~\cite{Yi:2021ryh,Wu:2022mkr,Jiang:2023vxp,Becattini:2019dxo,Becattini:2021suc,Becattini:2021iol,Liu:2020dxg,Liu:2021uhn,Fu:2021pok,Fu:2022myl}.

We assume the spin degree of freedom is in local thermal equilibrium on the freezeout hypersurface. The polarization vector of $\Lambda$ ($\bar{\Lambda}$) in its rest frame can then be obtained as
\begin{eqnarray}
\vec{P}^{*}(\vec{p}) = \vec{P}(\vec{p})-\frac{\vec{P}(\vec{p}) \cdot \vec{p}}{p^{0}(p^{0}+m)}\vec{p},
\end{eqnarray}
where $P^{\mu}(\vec{p}) \equiv  \mathcal{S}^{\mu}(\vec{p})/s$ with $s=1/2$ the particle spin.
After averaging over the transverse momentum, the local polarization is given by
\begin{eqnarray}
\vec{P}(\phi) = \frac{\int_{y_{\text{min}}}^{y_{\text{max}}}dy \int_{p_\text{T\text{min}}}^{p_\text{T\text{max}}}p_\text{T}dp_\text{T}
[ \Phi (\vec{p})\vec{P}^{*}(\vec{p})]}{\int_{y_{\text{min}}}^{y_{\text{max}}}dy \int_{p_\text{T\text{min}}}^{p_\text{T\text{max}}}p_\text{T}dp_\text{T} \Phi(\vec{p}) },
\label{eq:localP}
\end{eqnarray}
where $\phi$ is the azimuthal angle, and $\Phi(\vec{p})=\int d\Sigma\cdot p\, f_\mathrm{eq}$ with $f_\mathrm{eq}$ the equilibrium distribution of $\Lambda$ ($\bar{\Lambda}$) on the freezeout hypersurface. 
The mass of $\Lambda$ ($\bar{\Lambda}$) hyperon is set as $m = 1.116$~GeV. The global polarization of $\Lambda$ ($\bar{\Lambda}$) is obtained by further averaging $\vec{P}(\phi)$ over $\phi$. Contribution to $\Lambda$ ($\bar{\Lambda}$) polarization from resonance decays has not been included in our current calculation, which has been investigated in Refs.~\cite{Becattini:2016gvu,Palermo:2024tza} and can be implemented in our future work.

\begin{figure}[tbp!]
\includegraphics[width=0.85\linewidth]{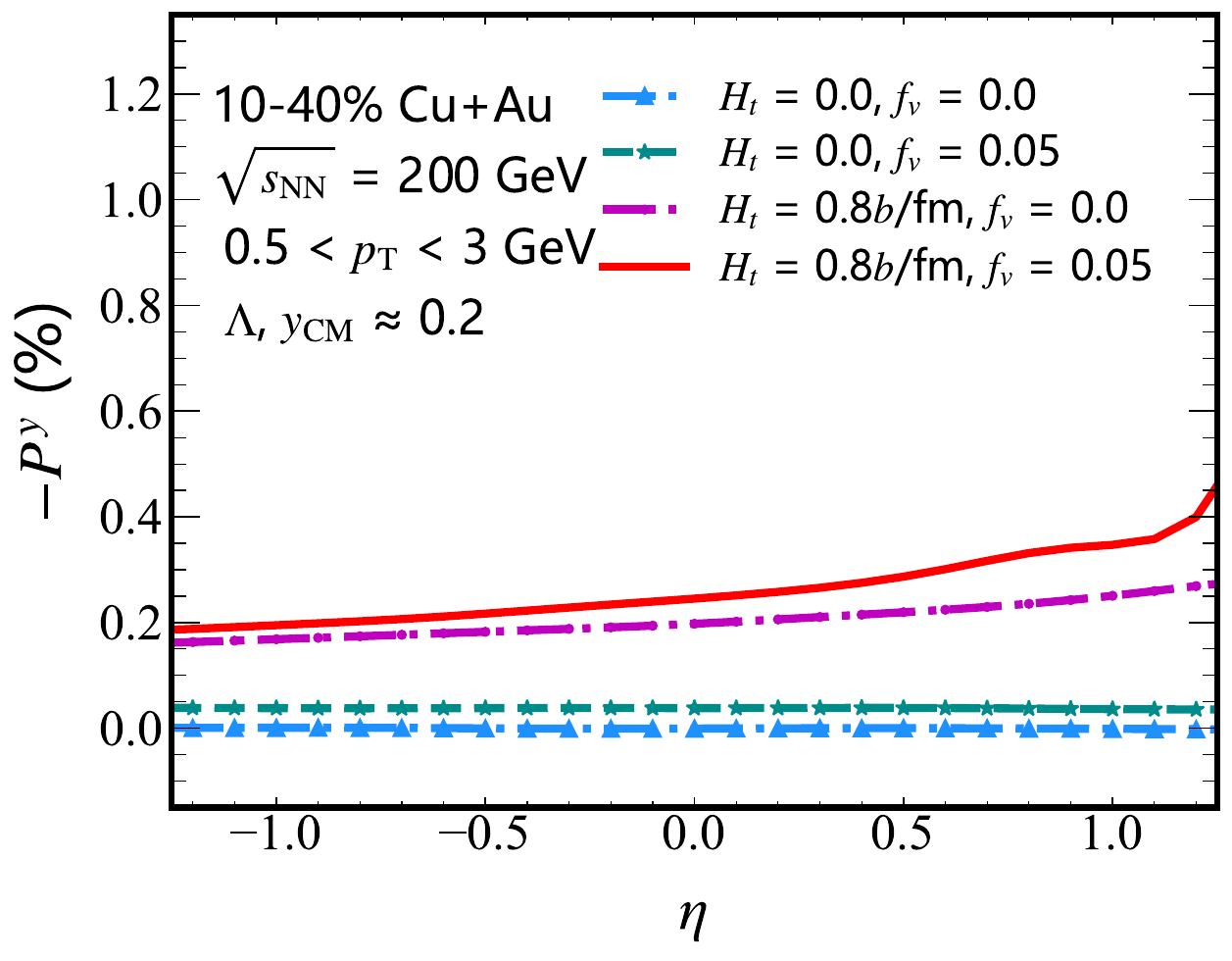} \\
\includegraphics[width=0.85\linewidth]{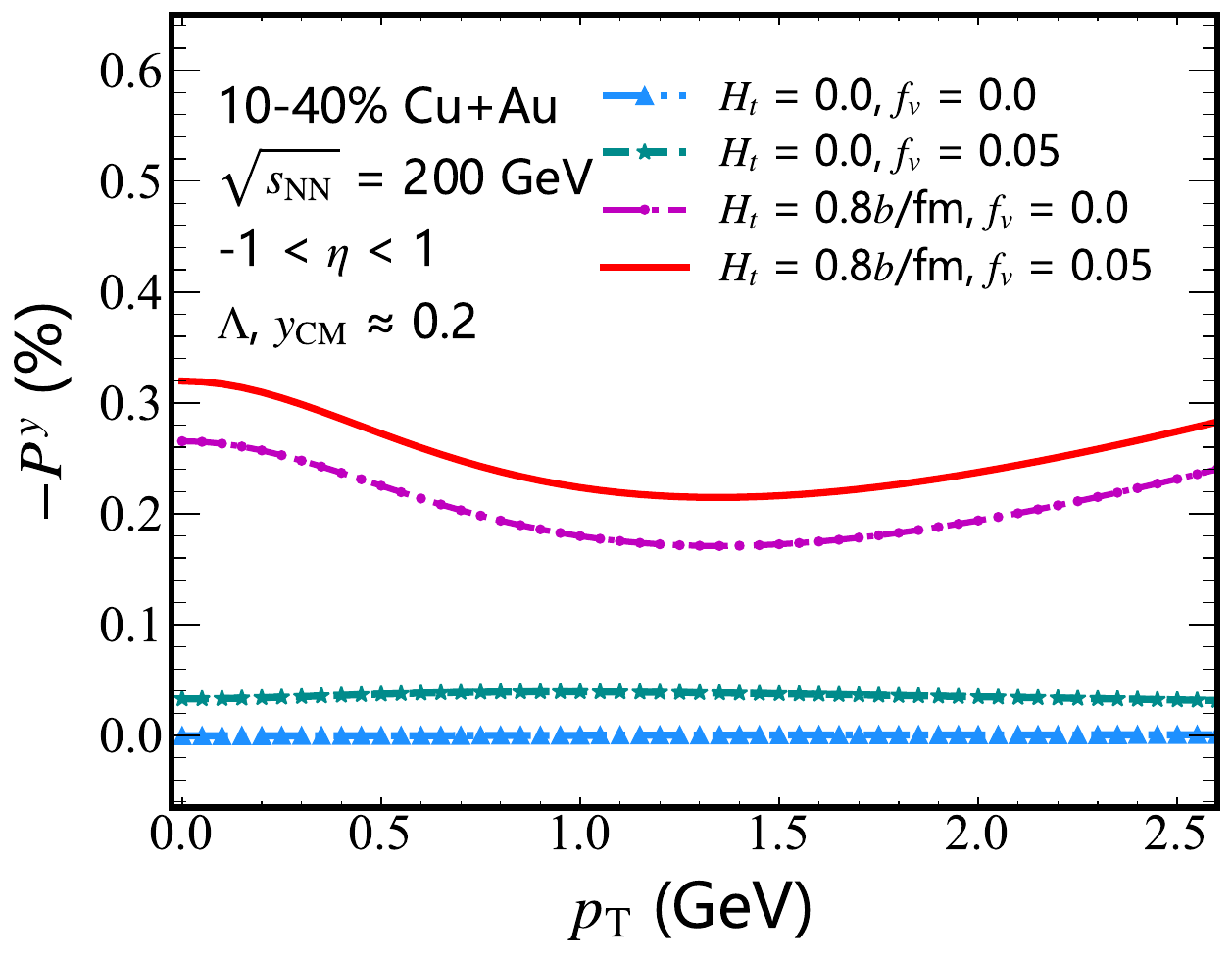}
\caption{(Color online) The global polarization of $\Lambda$ hyperon, $-P^{y}$, as functions of pseudorapidity (upper panel) and transverse momentum (lower panel) in Cu+Au collisions at $\snn=200$~GeV, compared between different initial conditions.}
\label{f:global_py_is}
\end{figure}

Using parameters fit from the directed flow coefficient in the previous subsection, $H_\text{t}=0.8b$/fm and $f_v=0.05$, we present in Fig.~\ref{f:global_py} our prediction for the global polarization of $\Lambda$ and $\bar{\Lambda}$ in Cu+Au collisions at $\snn=200$~GeV. Since the global polarization aligns with the orbital angular momentum of the heavy-ion collision system, i.e., $-\hat{y}$ directions, we show $-P^y$ with positive values here. In the left panel, we see the magnitude of the hyperon polarization increases with centrality, which can be understood as the larger orbital angular momentum deposited into the QGP as the impact parameter of heavy-ion collisions increases. In addition, we observe the magnitude of global polarization here in Cu+Au collisions is larger that in Au+Au collisions shown in our earlier calculation~\cite{Jiang:2023fai} and experimental data~\cite{STAR:2018gyt}. Within the same centrality bin, the QGP produced in Au+Au collisions has larger size than that in Cu+Au collisions, and thus the former expands for a longer time, leading to a larger inertial and therefore a smaller local vorticity in the end. This is consistent with the finding in Ref.~\cite{Shi:2017wpk}.

In the middle panel of Fig.~\ref{f:global_py}, we present the global polarization as a function of pseudorapidity. Different from Au+Au collisions in which $-P^y$ is an even function of $\eta$~\cite{Jiang:2023fai,Jiang:2023vxp}, the asymmetric energy deposition from Cu and Au nuclei generates larger $-P^{y}$ at forward rapidity (Cu-moving direction) than at backward rapidity (Au-moving direction). 

In the right panel of Fig.~\ref{f:global_py}, we show the global polarization as a function of transverse momentum. As discussed in our earlier study~\cite{Jiang:2023vxp}, the non-monotonic dependence of $-P^{y}$ on $p_\text{T}$ results from the competition between the thermal vorticity term that decreases with $p_\text{T}$ and the shear induced term that increases with $p_\text{T}$, in the presence of the initial tilt of the QGP. Compared to our earlier results for Au+Au collisions at $\snn=200$~GeV~\cite{Jiang:2023fai}, this non-monotonic behavior is more prominent here in asymmetric Cu+Au collisions. Note that within our calculation, the small difference in $-P^{y}$ between $\Lambda$ and $\bar{\Lambda}$ comes from the net baryon density distribution inside the QGP. The separation may be enhanced by considering effects of electromagnetic field produced in relativistic heavy-ion collisions~\cite{Guo:2019joy,Peng:2022cya,Xu:2022hql}.

In the end, we demonstrate effects of the tilted geometry and longitudinal flow velocity in our initial condition on the global polarization of hyperons. In the upper panel of Fig.~\ref{f:global_py_is}, we show the $\eta$ dependence of $-P^y$ and find that when $H_\text{t}=0$ and $f_{v}=0$, the global polarization $-P^{y}$ is zero. After introducing the tilted geometry of the QGP ($H_\text{t}\ne 0$), non-zero $-P^{y}$ can be generated during the hydrodynamic expansion of the QGP. As discussed earlier, the imbalanced energy deposition from Cu and Au nuclei causes the asymmetry of hyperon polarization between forward and backward rapidities. Incorporation of the initial longitudinal flow velocity ($f_v\ne 0$) further enhances the hyperon polarization, since this generates a positive gradient of the longitudinal fluid velocity along the $+\hat{x}$ direction, which increases the local vorticity of the QGP along the $-\hat{y}$ direction. Similarly, in the lower panel of Fig.~\ref{f:global_py_is} for the $p_\mathrm{T}$ dependence of $-P^y$, we see zero global polarization with $H_\text{t}=0$ and $f_{v}=0$. The initial tilted geometry of the QGP generates a considerable amount of hyperon polarization, which first decreases and then increases with $p_\mathrm{T}$. Further inclusion of the initial longitudinal flow gradient brings additional enhancement of the hyperon polarization. We note that both the overall magnitude and the non-monotonic $p_\mathrm{T}$ dependence of hyperon polarization is stronger in asymmetric Cu+Au collisions than in symmetric Au+Au collisions. Therefore, related measurements in Cu+Au collisions would provide a valuable opportunity for improving our understanding on the formation and evolution mechanisms of spin polarization in relativistic heavy-ion collisions, and its dependence on the initial condition of the QGP.

%%%%%%%%%%%%%%%%%%%%%%% Sec - Conclusion

\section{Conclusions}
\label{section4}

We have studied the collective flow coefficients and global polarization of hyperons in asymmetric Cu+Au collisions using an extended TRENTo initial condition model coupled to the (3+1)-D CLVisc hydrodynamic model. The 2D TRENTo initial condition has been extended to the 3D space, with a tilted geometry of the QGP medium with respect to the longitudinal direction and an initial longitudinal flow velocity profile introduced. Within this model, we are able to provide a satisfactory description of not only the $p_\mathrm{T}$ spectra of identified hadrons across various centrality bins, but also the $\eta$, $p_\mathrm{T}$ and centrality dependences of the charged hadron $v_1$ in Cu+Au collisions at $\snn=200$~GeV for the first time. Predictions on the hadron $v_2$ and global polarization of $\Lambda$ and $\bar{\Lambda}$ hyperons are also provided.

Effects of different components of our initial condition model on the collective flow coefficients and hyperon polarization have been systematically explored. We have found that the asymmetric energy deposition from Cu and Au nuclei between the $-\hat{x}$ and $+\hat{x}$ directions leads to positive $\varepsilon_1$ and then positive $v_1$ through hydrodynamic expansion in the transverse plane. The magnitudes of $\varepsilon_1$ and $v_1$ are significantly enhanced when the initial tilted geometry of the QGP is introduced. A counterclockwise tilt of the QGP in the participant plane with respect to the longitudinal direction makes $\varepsilon_1$ increase from negative to positive values as $\eta_\mathrm{s}$ increases, but $v_1$ decreases from positive to negative values as $\eta$ increases, through hydrodynamic expansion in the 3D space. In the presence of the tilted geometry, the charged particle $v_1$ exhibits clear positive values at low $p_\mathrm{T}$ but negative values at high $p_\mathrm{T}$ due to momentum conservation in the transverse plane. While the initial geometry determines the qualitative pattern of $v_1$ as functions of $\eta$ and $p_\mathrm{T}$, the initial longitudinal flow field introduces additional modification on $v_1$: it increases the values of $v_1$ and shifts the crossing points of $v_1(\eta)=0$ and $v_1(p_\mathrm{T})=0$ towards larger $\eta$ and $p_\mathrm{T}$ regions, respectively. We have also found the initial tilt of the QGP can result in a double peak structure of the charged hadron $v_2$ at forward and backward rapidities, and a non-monotonic $p_\mathrm{T}$ dependence of the global polarization of hyperons. The initial longitudinal flow field has no visible effect on the hadron $v_2$, but enhances the local vorticity of the QGP and therefore the hyperon polarization. Due to the imbalanced energy deposition from Cu and Au nuclei between forward and backward directions, a center-of-mass shift towards the Au-moving direction, which is stronger in more central collisions, is required when we compare results from hydrodynamic simulation and experimental data. The imbalanced energy deposition gives rise to an asymmetry between forward and backward rapidity regions for both collective flow coefficients and hyperon polarization. 

We note that the magnitudes of hadron $v_1$ and hyperon polarization are both larger in Cu+Au collisions than in Au+Au collisions due to different of geometries and sizes between these two systems. Characteristic signals of the tilted geometry, such as the double peak structure of $v_2$ and the non-monotonic $p_\mathrm{T}$ dependence of global polarization, are also stronger in Cu+Au than in Au+Au collisions. Therefore, Cu+Au collisions offer a valuable environment for investigating the production and evolution mechanisms of collective flows and spin polarization, as well as their dependences on the initial condition of the QGP. Since different observables -- $v_1$, $v_2$, and hyperon polarization -- have different sensitivities to the initial geometry and longitudinal flow field of the medium, a combined measurement on these observables would provide a tight constraint on the initial condition of the QGP in asymmetric heavy-ion collisions.
\\
\begin{acknowledgements}
We thank Xiang-Yu Wu, Jie Zhu, Cong Yi, Weiyao Ke, Zhenyu Chen, Hao-Jie Xu, Xiaofeng Luo, Li Yan, and Guang-You Qin for helpful discussions. 
This work was supported by the National Natural Science Foundation of China (NSFC) under Grant Nos.~12305138, 12175122, 2021-867, 11935007, and the
Guangdong Major Project of Basic and Applied Basic Research No.~2020B0301030008. 
\end{acknowledgements}

\bibliographystyle{unsrt}
\bibliography{clv3}

\begin{thebibliography}{10}

\bibitem{Busza:2018rrf}
Wit Busza, Krishna Rajagopal, and Wilke van~der Schee.
\newblock {Heavy Ion Collisions: The Big Picture, and the Big Questions}.
\newblock {\em Ann. Rev. Nucl. Part. Sci.}, 68:339--376, 2018.

\bibitem{STAR:2003wqp}
J.~Adams et~al.
\newblock {Particle type dependence of azimuthal anisotropy and nuclear modification of particle production in Au + Au collisions at s(NN)**(1/2) = 200-GeV}.
\newblock {\em Phys. Rev. Lett.}, 92:052302, 2004.

\bibitem{ALICE:2010suc}
K~Aamodt et~al.
\newblock {Elliptic flow of charged particles in Pb-Pb collisions at 2.76 TeV}.
\newblock {\em Phys. Rev. Lett.}, 105:252302, 2010.

\bibitem{ATLAS:2014txd}
Georges Aad et~al.
\newblock {Measurement of the centrality and pseudorapidity dependence of the integrated elliptic flow in lead-lead collisions at $\sqrt{s_{\mathrm {NN}}}=2.76$ TeV with the ATLAS detector}.
\newblock {\em Eur. Phys. J. C}, 74(8):2982, 2014.

\bibitem{CMS:2012zex}
Serguei Chatrchyan et~al.
\newblock {Measurement of the Elliptic Anisotropy of Charged Particles Produced in PbPb Collisions at $\sqrt{s}_{NN}$=2.76 TeV}.
\newblock {\em Phys. Rev. C}, 87(1):014902, 2013.

\bibitem{Voloshin:2008dg}
Sergei~A. Voloshin, Arthur~M. Poskanzer, and Raimond Snellings.
\newblock {Collective phenomena in non-central nuclear collisions}.
\newblock {\em Landolt-Bornstein}, 23:293--333, 2010.

\bibitem{Bilandzic:2010jr}
Ante Bilandzic, Raimond Snellings, and Sergei Voloshin.
\newblock {Flow analysis with cumulants: Direct calculations}.
\newblock {\em Phys. Rev. C}, 83:044913, 2011.

\bibitem{Huovinen:2001cy}
P.~Huovinen, P.~F. Kolb, Ulrich~W. Heinz, P.~V. Ruuskanen, and S.~A. Voloshin.
\newblock {Radial and elliptic flow at RHIC: Further predictions}.
\newblock {\em Phys. Lett. B}, 503:58--64, 2001.

\bibitem{Kolb:2003dz}
Peter~F. Kolb and Ulrich~W. Heinz.
\newblock {Hydrodynamic description of ultrarelativistic heavy ion collisions}.
\newblock pages 634--714, 5 2003.

\bibitem{Gale:2013da}
Charles Gale, Sangyong Jeon, and Bjoern Schenke.
\newblock {Hydrodynamic Modeling of Heavy-Ion Collisions}.
\newblock {\em Int. J. Mod. Phys. A}, 28:1340011, 2013.

\bibitem{Shen:2020mgh}
Chun Shen and Li~Yan.
\newblock {Recent development of hydrodynamic modeling in heavy-ion collisions}.
\newblock {\em Nucl. Sci. Tech.}, 31(12):122, 10 2020.

\bibitem{Policastro:2001yc}
G.~Policastro, Dan~T. Son, and Andrei~O. Starinets.
\newblock {The Shear viscosity of strongly coupled N=4 supersymmetric Yang-Mills plasma}.
\newblock {\em Phys. Rev. Lett.}, 87:081601, 2001.

\bibitem{Kovtun:2004de}
P.~Kovtun, Dan~T. Son, and Andrei~O. Starinets.
\newblock {Viscosity in strongly interacting quantum field theories from black hole physics}.
\newblock {\em Phys. Rev. Lett.}, 94:111601, 2005.

\bibitem{Bernhard:2019bmu}
Jonah~E. Bernhard, J.~Scott Moreland, and Steffen~A. Bass.
\newblock {Bayesian estimation of the specific shear and bulk viscosity of quark\textendash{}gluon plasma}.
\newblock {\em Nature Phys.}, 15(11):1113--1117, 2019.

\bibitem{JETSCAPE:2020shq}
D.~Everett et~al.
\newblock {Phenomenological constraints on the transport properties of QCD matter with data-driven model averaging}.
\newblock {\em Phys. Rev. Lett.}, 126(24):242301, 2021.

\bibitem{Schenke:2011bn}
Bjorn Schenke, Sangyong Jeon, and Charles Gale.
\newblock {Higher flow harmonics from (3+1)D event-by-event viscous hydrodynamics}.
\newblock {\em Phys. Rev. C}, 85:024901, 2012.

\bibitem{Gale:2012rq}
Charles Gale, Sangyong Jeon, Bj\"orn Schenke, Prithwish Tribedy, and Raju Venugopalan.
\newblock {Event-by-event anisotropic flow in heavy-ion collisions from combined Yang-Mills and viscous fluid dynamics}.
\newblock {\em Phys. Rev. Lett.}, 110(1):012302, 2013.

\bibitem{Heinz:2013th}
U.~Heinz and R.~Snellings.
\newblock {Collective flow and viscosity in relativistic heavy-ion collisions}.
\newblock {\em Ann. Rev. Nucl. Part. Sci.}, 63:123--151, 2013.

\bibitem{Chen:2024xbi}
Xinrong Chen, Xiang-Yu Wu, Shanshan Cao, and Guang-You Qin.
\newblock {System-size and shape dependencies of collective-flow fluctuations in relativistic nuclear collisions}.
\newblock {\em Phys. Rev. C}, 109(6):064915, 2024.

\bibitem{Zhang:2021kxj}
Chunjian Zhang and Jiangyong Jia.
\newblock {Evidence of Quadrupole and Octupole Deformations in Zr96+Zr96 and Ru96+Ru96 Collisions at Ultrarelativistic Energies}.
\newblock {\em Phys. Rev. Lett.}, 128(2):022301, 2022.

\bibitem{Jia:2021qyu}
Jiangyong Jia.
\newblock {Probing triaxial deformation of atomic nuclei in high-energy heavy ion collisions}.
\newblock {\em Phys. Rev. C}, 105(4):044905, 2022.

\bibitem{Bally:2022vgo}
Benjamin Bally et~al.
\newblock {Imaging the initial condition of heavy-ion collisions and nuclear structure across the nuclide chart}.
\newblock 9 2022.

\bibitem{Giacalone:2021clp}
Giuliano Giacalone, Bj\"orn Schenke, and Chun Shen.
\newblock {Constraining the Nucleon Size with Relativistic Nuclear Collisions}.
\newblock {\em Phys. Rev. Lett.}, 128(4):042301, 2022.

\bibitem{Zhu:2024tns}
Jie Zhu, Xiang-Yu Wu, and Guang-You Qin.
\newblock {Anisotropic flow, flow fluctuation and flow decorrelation in relativistic heavy-ion collisions: the roles of sub-nucleon structure and shear viscosity}.
\newblock {\em arXiv: 2401.15536}, 1 2024.

\bibitem{Bozek:2010bi}
P.~Bozek and I.~Wyskiel.
\newblock {Directed flow in ultrarelativistic heavy-ion collisions}.
\newblock {\em Phys. Rev. C}, 81:054902, 2010.

\bibitem{Jiang:2021ajc}
Ze-Fang Jiang, Shanshan Cao, Xiang-Yu Wu, C.~B. Yang, and Ben-Wei Zhang.
\newblock {Longitudinal distribution of initial energy density and directed flow of charged particles in relativistic heavy-ion collisions}.
\newblock {\em Phys. Rev. C}, 105(3):034901, 2022.

\bibitem{Bozek:2022svy}
Piotr Bozek.
\newblock {Splitting of proton-antiproton directed flow in relativistic heavy-ion collisions}.
\newblock {\em Phys. Rev. C}, 106(6):L061901, 2022.

\bibitem{Parida:2022zse}
Tribhuban Parida and Sandeep Chatterjee.
\newblock {Directed flow of light flavor hadrons for Au+Au collisions at $\sqrt{S_{NN}}=$ 7.7-200 GeV}.
\newblock {\em arXiv: 2211.15659}, 11 2022.

\bibitem{Jiang:2023fad}
Ze-Fang Jiang, Xiang-Yu Wu, Shanshan Cao, and Ben-Wei Zhang.
\newblock {Directed flow and global polarization in Au+Au collisions across energies covered by the beam energy scan at RHIC}.
\newblock {\em Phys. Rev. C}, 107(3):034904, 2023.

\bibitem{Shen:2020jwv}
Chun Shen and S.~Alzhrani.
\newblock {Collision-geometry-based 3D initial condition for relativistic heavy-ion collisions}.
\newblock {\em Phys. Rev. C}, 102(1):014909, 2020.

\bibitem{Ryu:2021lnx}
Sangwook Ryu, Vahidin Jupic, and Chun Shen.
\newblock {Probing early-time longitudinal dynamics with the \ensuremath{\Lambda} hyperon's spin polarization in relativistic heavy-ion collisions}.
\newblock {\em Phys. Rev. C}, 104(5):054908, 2021.

\bibitem{Jiang:2023vxp}
Ze-Fang Jiang, Xiang-Yu Wu, Shanshan Cao, and Ben-Wei Zhang.
\newblock {Hyperon polarization and its relation with directed flow in high-energy nuclear collisions}.
\newblock {\em Phys. Rev. C}, 108(6):064904, 2023.

\bibitem{Csernai:2011gg}
L.~P. Csernai, V.~K. Magas, H.~Stocker, and D.~D. Strottman.
\newblock {Fluid Dynamical Prediction of Changed v1-flow at LHC}.
\newblock {\em Phys. Rev. C}, 84:024914, 2011.

\bibitem{Retinskaya:2012ky}
Ekaterina Retinskaya, Matthew Luzum, and Jean-Yves Ollitrault.
\newblock {Directed flow at midrapidity in $\sqrt{s_{NN}}=2.76$ TeV Pb+Pb collisions}.
\newblock {\em Phys. Rev. Lett.}, 108:252302, 2012.

\bibitem{Ivanov:2016sqy}
Yu.~B. Ivanov and A.~A. Soldatov.
\newblock {What can we learn from the directed flow in heavy-ion collisions at BES RHIC energies?}
\newblock {\em Eur. Phys. J. A}, 52(1):10, 2016.

\bibitem{Nara:2016hbg}
Yasushi Nara, Harri Niemi, Jan Steinheimer, and Horst St\"ocker.
\newblock {Equation of state dependence of directed flow in a microscopic transport model}.
\newblock {\em Phys. Lett. B}, 769:543--548, 2017.

\bibitem{Du:2022yok}
Lipei Du, Chun Shen, Sangyong Jeon, and Charles Gale.
\newblock {Probing initial baryon stopping and equation~of state with rapidity-dependent directed flow of identified particles}.
\newblock {\em Phys. Rev. C}, 108(4):L041901, 2023.

\bibitem{Nayak:2023ofv}
Kishora Nayak, Shusu Shi, and Zi-Wei Lin.
\newblock {Coalescence sum rule and the electric charge- and strangeness-dependences of directed flow in heavy ion collisions}.
\newblock {\em Phys. Lett. B}, 849:138479, 2024.

\bibitem{Selyuzhenkov:2011zj}
Ilya Selyuzhenkov.
\newblock {Charged particle directed flow in Pb-Pb collisions at $\sqrt{s_{NN}}=2.76$ TeV measured with ALICE at the LHC}.
\newblock {\em J. Phys. G}, 38:124167, 2011.

\bibitem{ATLAS:2012at}
Georges Aad et~al.
\newblock {Measurement of the azimuthal anisotropy for charged particle production in $\sqrt{s_{NN}}=2.76$ TeV lead-lead collisions with the ATLAS detector}.
\newblock {\em Phys. Rev. C}, 86:014907, 2012.

\bibitem{STAR:2016cio}
L.~Adamczyk et~al.
\newblock {Charge-dependent directed flow in Cu+Au collisions at $\sqrt{s_{_{NN}}}$ = 200 GeV}.
\newblock {\em Phys. Rev. Lett.}, 118(1):012301, 2017.

\bibitem{STAR:2022fnj}
Bassam Aboona et~al.
\newblock {Observation of Directed Flow of Hypernuclei H\ensuremath{\Lambda}3 and H\ensuremath{\Lambda}4 in sNN=3\,\,GeV Au+Au Collisions at RHIC}.
\newblock {\em Phys. Rev. Lett.}, 130(21):212301, 2023.

\bibitem{ALICE:2013xri}
Betty Abelev et~al.
\newblock {Directed Flow of Charged Particles at Midrapidity Relative to the Spectator Plane in Pb-Pb Collisions at $\sqrt{s_{NN}}$=2.76 TeV}.
\newblock {\em Phys. Rev. Lett.}, 111(23):232302, 2013.

\bibitem{STAR:2017okv}
L.~Adamczyk et~al.
\newblock {Beam-Energy Dependence of Directed Flow of $\Lambda$, $\bar{\Lambda}$, $K^\pm$, $K^0_s$ and $\phi$ in Au+Au Collisions}.
\newblock {\em Phys. Rev. Lett.}, 120(6):062301, 2018.

\bibitem{STAR:2023jdd}
M.~I. Abdulhamid et~al.
\newblock {Observation of the electromagnetic field effect via charge-dependent directed flow in heavy-ion collisions at the Relativistic Heavy Ion Collider}.
\newblock {\em Phys. Rev. X}, 14(1):011028, 2024.

\bibitem{STAR:2024ujm}
{Measurement of directed flow in Au+Au collisions at $\sqrt{s_{NN}}=$ 19.6 and 27 GeV with the STAR Event Plane Detector}.
\newblock 6 2024.

\bibitem{STAR:2017ykf}
Leszek Adamczyk et~al.
\newblock {Azimuthal anisotropy in Cu$+$Au collisions at $\sqrt{s_{_{NN}}}$ = 200 GeV}.
\newblock {\em Phys. Rev. C}, 98(1):014915, 2018.

\bibitem{Bozek:2012hy}
Piotr Bo\.zek.
\newblock {Event-by-event viscous hydrodynamics for Cu\textendash{}Au collisions at $\sqrt{s_{NN}}$=200 GeV}.
\newblock {\em Phys. Lett. B}, 717:287--290, 2012.

\bibitem{Nakamura:2022idq}
Kouki Nakamura, Takahiro Miyoshi, Chiho Nonaka, and Hiroyuki~R. Takahashi.
\newblock {Directed flow in relativistic resistive magneto-hydrodynamic expansion for symmetric and asymmetric collision systems}.
\newblock {\em Phys. Rev. C}, 107(1):014901, 2023.

\bibitem{Chen:2005zy}
Lie-Wen Chen and Che~Ming Ko.
\newblock {Anisotropic flow in Cu+Au collisions at $\sqrt{s_{NN}}$ = 200 GeV}.
\newblock {\em Phys. Rev. C}, 73:014906, 2006.

\bibitem{Voronyuk:2014rna}
V.~Voronyuk, V.~D. Toneev, S.~A. Voloshin, and W.~Cassing.
\newblock {Charge-dependent directed flow in asymmetric nuclear collisions}.
\newblock {\em Phys. Rev. C}, 90(6):064903, 2014.

\bibitem{Moreland:2014oya}
J.~Scott Moreland, Jonah~E. Bernhard, and Steffen~A. Bass.
\newblock {Alternative ansatz to wounded nucleon and binary collision scaling in high-energy nuclear collisions}.
\newblock {\em Phys. Rev. C}, 92(1):011901, 2015.

\bibitem{Pang:2018zzo}
Long-Gang Pang, H.~Petersen, and Xin-Nian Wang.
\newblock {Pseudorapidity distribution and decorrelation of anisotropic flow within the open-computing-language implementation CLVisc hydrodynamics}.
\newblock {\em Phys. Rev. C}, 97(6):064918, 2018.

\bibitem{Wu:2021fjf}
Xiang-Yu Wu, Guang-You Qin, Long-Gang Pang, and Xin-Nian Wang.
\newblock {(3+1)-D viscous hydrodynamics at finite net baryon density: Identified particle spectra, anisotropic flows, and flow fluctuations across energies relevant to the beam-energy scan at RHIC}.
\newblock {\em Phys. Rev. C}, 105(3):034909, 2022.

\bibitem{Bernhard:2016tnd}
Jonah~E. Bernhard, J.~Scott Moreland, Steffen~A. Bass, Jia Liu, and Ulrich Heinz.
\newblock {Applying Bayesian parameter estimation to relativistic heavy-ion collisions: simultaneous characterization of the initial state and quark-gluon plasma medium}.
\newblock {\em Phys. Rev. C}, 94(2):024907, 2016.

\bibitem{trento:2d}
Jonah~E. Bernhard, J.~Scott Moreland, and Steffen~A. Bass.
\newblock https://github.com/duke-qcd/trento.
\newblock 2019.
\newblock \url{https://github.com/Duke-QCD/trento}.

\bibitem{Alver:2008aq}
B.~Alver, M.~Baker, C.~Loizides, and P.~Steinberg.
\newblock {The PHOBOS Glauber Monte Carlo}.
\newblock 5 2008.

\bibitem{Loizides:2017ack}
C.~Loizides, J.~Kamin, and D.~d'Enterria.
\newblock {Improved Monte Carlo Glauber predictions at present and future nuclear colliders}.
\newblock {\em Phys. Rev. C}, 97(5):054910, 2018.
\newblock [Erratum: Phys.Rev.C 99, 019901 (2019)].

\bibitem{Jiang:2021foj}
Ze-Fang Jiang, C.~B. Yang, and Qi~Peng.
\newblock {Directed flow of charged particles within idealized viscous hydrodynamics at energies available at the BNL Relativistic Heavy Ion Collider and at the CERN Large Hadron Collider}.
\newblock {\em Phys. Rev. C}, 104(6):064903, 2021.

\bibitem{Bozek:2011ua}
Piotr Bozek.
\newblock {Flow and interferometry in 3+1 dimensional viscous hydrodynamics}.
\newblock {\em Phys. Rev. C}, 85:034901, 2012.

\bibitem{Bozek:2016kpf}
Piotr Bo\.zek, Wojciech Broniowski, and Maciej Rybczy\'nski.
\newblock {Wounded quarks in A+A, p+A, and p+p collisions}.
\newblock {\em Phys. Rev. C}, 94(1):014902, 2016.

\bibitem{Denicol:2018wdp}
Gabriel~S. Denicol, Charles Gale, Sangyong Jeon, Akihiko Monnai, Bj\"orn Schenke, and Chun Shen.
\newblock {Net baryon diffusion in fluid dynamic simulations of relativistic heavy-ion collisions}.
\newblock {\em Phys. Rev. C}, 98(3):034916, 2018.

\bibitem{Alzhrani:2022dpi}
Sahr Alzhrani, Sangwook Ryu, and Chun Shen.
\newblock {\ensuremath{\Lambda} spin polarization in event-by-event relativistic heavy-ion collisions}.
\newblock {\em Phys. Rev. C}, 106(1):014905, 2022.

\bibitem{Qiu:2011iv}
Zhi Qiu and Ulrich~W. Heinz.
\newblock {Event-by-event shape and flow fluctuations of relativistic heavy-ion collision fireballs}.
\newblock {\em Phys. Rev. C}, 84:024911, 2011.

\bibitem{Monnai:2019hkn}
Akihiko Monnai, Bj\"orn Schenke, and Chun Shen.
\newblock {Equation of state at finite densities for QCD matter in nuclear collisions}.
\newblock {\em Phys. Rev. C}, 100(2):024907, 2019.

\bibitem{Monnai:2021kgu}
Akihiko Monnai, Bj\"orn Schenke, and Chun Shen.
\newblock {QCD Equation of State at Finite Chemical Potentials for Relativistic Nuclear Collisions}.
\newblock {\em Int. J. Mod. Phys. A}, 36(07):2130007, 2021.

\bibitem{PHENIX:2023kax}
N.~J. Abdulameer et~al.
\newblock {Identified charged-hadron production in p+Al, He3+Au, and Cu+Au collisions at sNN=200GeV and in U+U collisions at sNN=193GeV}.
\newblock {\em Phys. Rev. C}, 109(5):054910, 2024.

\bibitem{Teaney:2010vd}
Derek Teaney and Li~Yan.
\newblock {Triangularity and Dipole Asymmetry in Heavy Ion Collisions}.
\newblock {\em Phys. Rev. C}, 83:064904, 2011.

\bibitem{PHENIX:2015zbc}
A.~Adare et~al.
\newblock {Measurements of directed, elliptic, and triangular flow in Cu$+$Au collisions at $\sqrt{s_{_{NN}}}=200$ GeV}.
\newblock {\em Phys. Rev. C}, 94(5):054910, 2016.

\bibitem{PHOBOS:2004vcu}
B.~B. Back et~al.
\newblock {Centrality and pseudorapidity dependence of elliptic flow for charged hadrons in Au+Au collisions at s(NN)**(1/2) = 200-GeV}.
\newblock {\em Phys. Rev. C}, 72:051901, 2005.

\bibitem{ALICE:2016tlx}
Jaroslav Adam et~al.
\newblock {Pseudorapidity dependence of the anisotropic flow of charged particles in Pb-Pb collisions at $\sqrt{s_{\rm NN}}=2.76$ TeV}.
\newblock {\em Phys. Lett. B}, 762:376--388, 2016.

\bibitem{Voloshin:2017kqp}
Sergei~A. Voloshin.
\newblock {Vorticity and particle polarization in heavy ion collisions (experimental perspective)}.
\newblock {\em EPJ Web Conf.}, 171:07002, 2018.

\bibitem{Ivanov:2020wak}
Yu.~B. Ivanov and A.~A. Soldatov.
\newblock {Correlation between global polarization, angular momentum, and flow in heavy-ion collisions}.
\newblock {\em Phys. Rev. C}, 102(2):024916, 2020.

\bibitem{Yi:2021ryh}
Cong Yi, Shi Pu, and Di-Lun Yang.
\newblock {Reexamination of local spin polarization beyond global equilibrium in relativistic heavy ion collisions}.
\newblock {\em Phys. Rev. C}, 104(6):064901, 2021.

\bibitem{Wu:2022mkr}
Xiang-Yu Wu, Cong Yi, Guang-You Qin, and Shi Pu.
\newblock {Local and global polarization of \ensuremath{\Lambda} hyperons across RHIC-BES energies: The roles of spin hall effect, initial condition, and baryon diffusion}.
\newblock {\em Phys. Rev. C}, 105(6):064909, 2022.

\bibitem{Yi:2023tgg}
Cong Yi, Xiang-Yu Wu, Di-Lun Yang, Jian-Hua Gao, Shi Pu, and Guang-You Qin.
\newblock {Probing vortical structures in heavy-ion collisions at RHIC-BES energies through helicity polarization}.
\newblock {\em Phys. Rev. C}, 109(1):L011901, 2024.

\bibitem{Becattini:2013fla}
F.~Becattini, V.~Chandra, L.~Del~Zanna, and E.~Grossi.
\newblock {Relativistic distribution function for particles with spin at local thermodynamical equilibrium}.
\newblock {\em Annals Phys.}, 338:32--49, 2013.

\bibitem{Fang:2016vpj}
Ren-hong Fang, Long-gang Pang, Qun Wang, and Xin-nian Wang.
\newblock {Polarization of massive fermions in a vortical fluid}.
\newblock {\em Phys. Rev. C}, 94(2):024904, 2016.

\bibitem{Karpenko:2018erl}
Iurii Karpenko and Francesco Becattini.
\newblock {Lambda polarization in heavy ion collisions: from RHIC BES to LHC energies}.
\newblock {\em Nucl. Phys. A}, 982:519--522, 2019.

\bibitem{Hidaka:2017auj}
Yoshimasa Hidaka, Shi Pu, and Di-Lun Yang.
\newblock {Nonlinear Responses of Chiral Fluids from Kinetic Theory}.
\newblock {\em Phys. Rev. D}, 97(1):016004, 2018.

\bibitem{Yi:2021unq}
Cong Yi, Shi Pu, Jian-Hua Gao, and Di-Lun Yang.
\newblock {Hydrodynamic helicity polarization in relativistic heavy ion collisions}.
\newblock {\em Phys. Rev. C}, 105(4):044911, 2022.

\bibitem{Liu:2020dxg}
Shuai Y.~F. Liu and Yi~Yin.
\newblock {Spin Hall effect in heavy-ion collisions}.
\newblock {\em Phys. Rev. D}, 104(5):054043, 2021.

\bibitem{Becattini:2019dxo}
F.~Becattini, M.~Buzzegoli, and E.~Grossi.
\newblock {Reworking the Zubarev's approach to non-equilibrium quantum statistical mechanics}.
\newblock {\em Particles}, 2(2):197--207, 2019.

\bibitem{Becattini:2021suc}
F.~Becattini, M.~Buzzegoli, and A.~Palermo.
\newblock {Spin-thermal shear coupling in a relativistic fluid}.
\newblock {\em Phys. Lett. B}, 820:136519, 2021.

\bibitem{Becattini:2021iol}
F.~Becattini, M.~Buzzegoli, G.~Inghirami, I.~Karpenko, and A.~Palermo.
\newblock {Local Polarization and Isothermal Local Equilibrium in Relativistic Heavy Ion Collisions}.
\newblock {\em Phys. Rev. Lett.}, 127(27):272302, 2021.

\bibitem{Liu:2021uhn}
Shuai Y.~F. Liu and Yi~Yin.
\newblock {Spin polarization induced by the hydrodynamic gradients}.
\newblock {\em JHEP}, 07:188, 2021.

\bibitem{Fu:2021pok}
Baochi Fu, Shuai Y.~F. Liu, Longgang Pang, Huichao Song, and Yi~Yin.
\newblock {Shear-Induced Spin Polarization in Heavy-Ion Collisions}.
\newblock {\em Phys. Rev. Lett.}, 127(14):142301, 2021.

\bibitem{Fu:2022myl}
Baochi Fu, Longgang Pang, Huichao Song, and Yi~Yin.
\newblock {Signatures of the spin Hall effect in hot and dense QCD matter}.
\newblock {\em arXiv:2201.12970}, 2022.

\bibitem{Becattini:2016gvu}
F.~Becattini, I.~Karpenko, M.~Lisa, I.~Upsal, and S.~Voloshin.
\newblock {Global hyperon polarization at local thermodynamic equilibrium with vorticity, magnetic field and feed-down}.
\newblock {\em Phys. Rev. C}, 95(5):054902, 2017.

\bibitem{Palermo:2024tza}
Andrea Palermo, Eduardo Grossi, Iurii Karpenko, and Francesco Becattini.
\newblock {$\Lambda$ polarization in very high energy heavy ion collisions as a probe of the Quark-Gluon Plasma formation and properties}.
\newblock {\em arXiv: 2404.14295}, 4 2024.

\bibitem{Jiang:2023fai}
Ze-Fang Jiang, Xiang-Yu Wu, Hua-Qing Yu, Shan-Shan Cao, and Ben-Wei Zhang.
\newblock {The direct flow of charged particles and the global polarization of hyperons in 200 AGeV Au+Au collisions at RHIC}.
\newblock {\em Acta Phys. Sin.}, 72(7):072504, 2023.

\bibitem{STAR:2018gyt}
Jaroslav Adam et~al.
\newblock {Global polarization of $\Lambda$ hyperons in Au+Au collisions at $\sqrt{s_{_{NN}}}$ = 200 GeV}.
\newblock {\em Phys. Rev. C}, 98:014910, 2018.

\bibitem{Shi:2017wpk}
Shuzhe Shi, Kangle Li, and Jinfeng Liao.
\newblock {Searching for the Subatomic Swirls in the CuCu and CuAu Collisions}.
\newblock {\em Phys. Lett. B}, 788:409--413, 2019.

\bibitem{Guo:2019joy}
Yu~Guo, Shuzhe Shi, Shengqin Feng, and Jinfeng Liao.
\newblock {Magnetic Field Induced Polarization Difference between Hyperons and Anti-hyperons}.
\newblock {\em Phys. Lett. B}, 798:134929, 2019.

\bibitem{Peng:2022cya}
Hao-Hao Peng, Sihao Wu, Ren-jie Wang, Duan She, and Shi Pu.
\newblock {Anomalous magnetohydrodynamics with temperature-dependent electric conductivity and application to the global polarization}.
\newblock {\em Phys. Rev. D}, 107(9):096010, 2023.

\bibitem{Xu:2022hql}
Kun Xu, Fan Lin, Anping Huang, and Mei Huang.
\newblock {$\Lambda$/$\overline{\Lambda}$ polarization and splitting induced by rotation and magnetic field}.
\newblock {\em Phys. Rev. D}, 106(7):L071502, 2022.

\end{thebibliography}

\end{document}